\documentclass[twoside,twocolumn,9pt,floatfix,amsmath]{article}
\usepackage{extsizes}
\usepackage[super,sort&compress,comma]{natbib} 
\usepackage[version=4]{mhchem}
\usepackage[left=1.5cm, right=1.5cm, top=1.785cm, bottom=2.0cm]{geometry}
\usepackage{balance}
\usepackage{mathptmx}
\usepackage{sectsty}
\usepackage{graphicx} 
\usepackage{lastpage}
\usepackage[format=plain,justification=justified,singlelinecheck=false,font={stretch=1.125,small,sf},labelfont=bf,labelsep=space]{caption}
\usepackage{float}
\usepackage{fancyhdr}
\usepackage{fnpos}
\usepackage[english]{babel}
\addto{\captionsenglish}{%
  
}
\usepackage{array}
\usepackage{droidsans}
\usepackage{charter}
\usepackage[T1]{fontenc}
\usepackage[usenames,dvipsnames]{xcolor}
\usepackage{setspace}
\usepackage[compact]{titlesec}
\usepackage{hyperref}
\usepackage{gensymb}
\usepackage{amssymb,amsmath}
\usepackage{float}

\usepackage{epstopdf}

\definecolor{cream}{RGB}{222,217,201}

\begin{document}

\pagestyle{fancy}
\thispagestyle{plain}
\fancypagestyle{plain}{
\renewcommand{\headrulewidth}{0pt}
}

\makeFNbottom
\makeatletter
\renewcommand\LARGE{\@setfontsize\LARGE{15pt}{17}}
\renewcommand\Large{\@setfontsize\Large{12pt}{14}}
\renewcommand\large{\@setfontsize\large{10pt}{12}}
\renewcommand\footnotesize{\@setfontsize\footnotesize{7pt}{10}}
\makeatother

\renewcommand{\thefootnote}{\fnsymbol{footnote}}
\renewcommand\footnoterule{\vspace*{1pt}%
\color{cream}\hrule width 3.5in height 0.4pt \color{black}\vspace*{5pt}} 
\setcounter{secnumdepth}{5}

\makeatletter 
\renewcommand\@biblabel[1]{#1}            
\renewcommand\@makefntext[1]%
{\noindent\makebox[0pt][r]{\@thefnmark\,}#1}
\makeatother 
\renewcommand{\figurename}{\small{Fig.}~}
\sectionfont{\sffamily\Large}
\subsectionfont{\normalsize}
\subsubsectionfont{\bf}
\setstretch{1.125} 
\setlength{\skip\footins}{0.8cm}
\setlength{\footnotesep}{0.25cm}
\setlength{\jot}{10pt}
\titlespacing*{\section}{0pt}{4pt}{4pt}
\titlespacing*{\subsection}{0pt}{15pt}{1pt}

\fancyfoot{}
\fancyfoot[LO,RE]{\vspace{-7.1pt}\includegraphics[height=9pt]{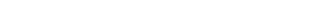}}
\fancyfoot[CO]{\vspace{-7.1pt}\hspace{11.9cm}\includegraphics{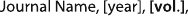}}
\fancyfoot[CE]{\vspace{-7.2pt}\hspace{-13.2cm}\includegraphics{head_foot/RF}}
\fancyfoot[RO]{\footnotesize{\sffamily{1--\pageref{LastPage} ~\textbar  \hspace{2pt}\thepage}}}
\fancyfoot[LE]{\footnotesize{\sffamily{\thepage~\textbar\hspace{4.65cm} 1--\pageref{LastPage}}}}
\fancyhead{}
\renewcommand{\headrulewidth}{0pt} 
\renewcommand{\footrulewidth}{0pt}
\setlength{\arrayrulewidth}{1pt}
\setlength{\columnsep}{6.5mm}
\setlength\bibsep{1pt}

\makeatletter 
\newlength{\figrulesep} 
\setlength{\figrulesep}{0.5\textfloatsep} 

\newcommand{\topfigrule}{\vspace*{-1pt}%
\noindent{\color{cream}\rule[-\figrulesep]{\columnwidth}{1.5pt}} }

\newcommand{\botfigrule}{\vspace*{-2pt}%
\noindent{\color{cream}\rule[\figrulesep]{\columnwidth}{1.5pt}} }

\newcommand{\dblfigrule}{\vspace*{-1pt}%
\noindent{\color{cream}\rule[-\figrulesep]{\textwidth}{1.5pt}} }

\makeatother

\twocolumn[
  \begin{@twocolumnfalse}
{\includegraphics[height=30pt]{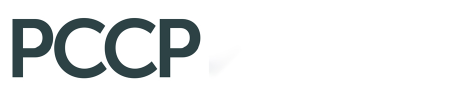}\hfill\raisebox{0pt}[0pt][0pt]{\includegraphics[height=55pt]{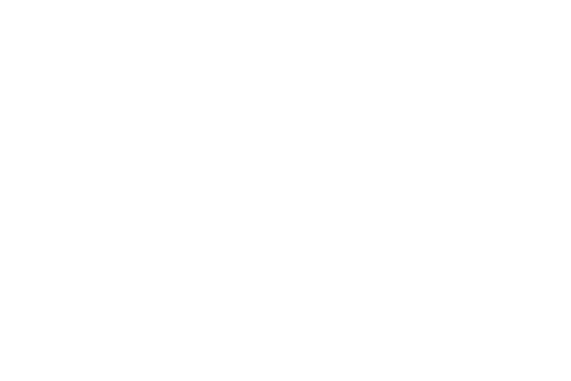}}\\[1ex]
\includegraphics[width=18.5cm]{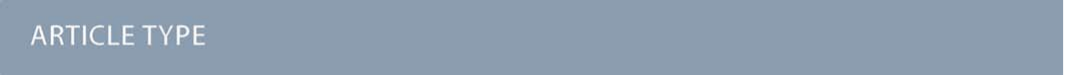}}\par
\vspace{1em}
\sffamily
\begin{tabular}{m{4.5cm} p{13.5cm} }

\includegraphics{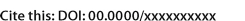} & \noindent\LARGE{\textbf{Ionic photofragmentation cross sections of the \ce{HS^+}, \ce{H_2S^+} and \ce{HCl^+} molecular ions near the $2p$ threshold}} \\
\vspace{0.3cm} & \vspace{0.3cm} \\

 & \noindent\large{Jean-Paul Mosnier,$^{\ast}$\textit{$^{a}$} Eugene T. Kennedy,\textit{$^{a}$} Denis Cubaynes,\textit{$^{b,c}$} Jean-Marc Bizau,\textit{$^{b,c}$} S\'{e}gol\`{e}ne Guilbaud,\textit{$^{c}$} and St\'{e}phane Carniato\textit{$^{d}$}} \\

\includegraphics{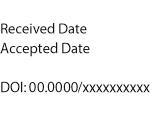} & \noindent\normalsize{The absolute cross sections for the production of X$^{2+}$ and X$^{3+}$ ions following absorption of monochromatised SOLEIL synchrotron radiation by the \ce{HX^+} hydride molecular ions (\ce{X} = \ce{S,Cl}) are presented as a function of photon energy in the region of the \ce{X_{2p}} ionisation thresholds ($\sim$ 180 eV and $\sim$ 220 eV for sulphur and chlorine, respectively). The experimental results are interpreted with the help of extensive ab initio density functional theory (DFT) and post-Hartree-Fock configuration interaction calculations including spin-orbit coupling to compute the absorption oscillator strengths of the \ce{X} 2p core excitations to valence and Rydberg states. In order to account for all the experimental features, the calculations must include vibrational dynamics and spin-orbit coupling. Similar experimental data are also presented for the sulfaniumyl \ce{H_2S^+} molecular ion.} \\


\end{tabular}

 \end{@twocolumnfalse} \vspace{0.6cm}

  ]

\renewcommand*\rmdefault{bch}\normalfont\upshape
\rmfamily
\section*{}
\vspace{-1cm}


\footnotetext{\textit{$^{\ast}$~Corresponding author; E-mail: jean-paul.mosnier@dcu.ie }}
\footnotetext{\textit{$^{a}~$School of Physical Sciences, Dublin City University, Dublin 9, Ireland.}}
\footnotetext{\textit{$^{b}$~Synchrotron SOLEIL, L'Orme des Merisiers, Saint-Aubin, BP 48, F-91192 Gif-sur-Yvette, CEDEX, France.}}
\footnotetext{\textit{$^{c}$~Institut des Sciences Mol\'{e}culaires d'Orsay, UMR 8214, Rue Andr\'{e} Rivi\`{e}re, B\^{a}timent 520, Universit\'{e} Paris-Saclay, F-91405 Orsay Cedex, France.}}
\footnotetext{\textit{$^{d}$~Laboratoire de Chimie Physique-Mati\`{e}re et Rayonnement (LCPMR), UMR 7614, Sorbonne Universit\'{e}, Campus Pierre et Marie Curie, 4 place Jussieu F-75005 Paris, France.}}








\section{\label{Int}Introduction}
The Auger decay of free atomic and molecular species upon the absorption of short wavelength (ionising) photons is a fundamental process in nature with relevance and applications in many different fields such as plasma physics \citep{Guha_2008}, astrophysics \citep{Adamkovics_2011,Gatuzz_2024}, medical physics \citep{Howell_2008} and the analytical, surface and material sciences, e.g. Ref.~\citep{Hofmann_2014}. Due to the strengthening of the Coulomb potential, the systematic study of the photoionisation behaviour of \emph{ionised} atomic and molecular species \citep{Schippers_2020, Kennedy_2022} is of equal fundamental interest. In the case of molecular ions, the non-spherical nature of the molecular field together with the coupling of the nuclear degrees of freedom will add complexity compared to the atomic case as seen in unique spectral features such as the molecular shape resonance \citep{Dehmer_1972}. 

While for atomic ions there exists a large corpus of experimental and theoretical photoionisation data, this is not the case for molecular ions. This is largely due to the significant challenges posed to both the experimental techniques and the theoretical models, namely (1) the production of sufficiently sensitive molecular ion beam or trap photoionisation apparatus and, (2) the accurate description of highly excited molecular states, respectively. In spite of these difficulties, the field of soft x-ray absorption in molecular ions has witnessed marked progress over the past ten years with combined experimental and theoretical modelling studies for, e.g., \ce{N_2^+} ~\citep{Lindblad_2020}, \ce{NO^+} ~\citep{Lindblad_2022}, \ce{H_3O^+} ~\citep{Schwarz_2022}, \ce{C_3H_3^+} ~\citep{Reinwardt_2024} or \ce{O_2^+} ~\citep{Cornetta_2025}. This is in parallel with a body of work on the structure and dynamics of molecular hydride ions: \ce{HC^+}, \ce{HO^+}, and \ce{HSi^+}~\citep{Mosnier_2016}, \ce{HI^+} ~\citep{Klumpp_2018}, \ce{HSi^+}, ~\ce{H_2Si^+}, \ce{H_3Si^+}\citep{Kennedy_2018}, \ce{H_yN^+} (\ce{y}= 0 to 3), ~\citep{Bari_2019}, \ce{HBr^+}~\citep{Kobayashi_2019}, \ce{HN^+} ~\citep{Carniato_2020} and \ce{HF^+} ~\citep{Martins_2021}. Further, the Auger decay rates and fragmentation dynamics of \ce{H_2C^+} and \ce{H_3C^+} were investigated theoretically in Ref.~\citep{Puglisi_2018}.

It is well known that diatomic molecular hydrides and their ions play key roles in the evolution of the interstellar medium and are, thus, invaluable probes (tracers) of many chemical and radiative processes in molecular clouds, e.g.~\citep{Walsh_2012, Abel_2008, Stauber_2005, Benz_2010, Goicoechea_2021, Gerin_2016}. In this work, we present for the first time experimental and theoretical absorption cross sections for the sulfanylium \ce{HS^+} and chloroniumyl \ce{HCl^+} diatomic molecular hydrides. These were discovered in the interstellar gas in 2011~\citep{Menten_2011} and 2012~\citep{De-Luca_2012}, respectively. We also report experimental cross sections for the sulfaniumyl \ce{H_2S^+} molecular ion which, to our knowledge, remains hitherto unidentified in the interstellar gas. 

Another point of interest concerning the inner shell photoionisation of molecular hydrides is that the initial core hole and its subsequent decay are necessarily localised on the heavy element. If the non-radiative decay leads to dissociation of the original parent molecular ion, the fragments are imparted additional kinetic energy-, ie the phenomenon of kinetic energy release or KER. As the KER mirrors the Auger decay~\citep{Chiang_2012}, related experimental and theoretical studies are key to the understanding of the interplay between electronic and nuclear decay processes in molecular photoionisation. This has been an extensively researched field for the past nearly sixty years and is still actively explored, see e.g. references~\citep{Carlson_1966, Hitchcock_1988, Nenner_1996, Ueda_2005, Chiang_2012, Falcinelli_2014, Scarlett_2017, Piancastelli_2017, Rebholz_2021, Martins_2021} and all the references therein.
 
 In this work, we report absolute photofragmentation cross sections for the \ce{HS^+}, \ce{H_2S^+} and \ce{HCl^+} molecular ions in the photon energy regions of the $2p$ threshold. The specific experimental aspects of our photoion yield measurements are detailed in the experimental section. This is supplemented with a discussion (in Appendix), based on an analytical approach, of some physical aspects of KER that are pertinent to our cross section measurements carried out using a fast ion beam. In Section~\ref{Results} Analyses and discussions, the experimental results are analysed with the help of ab initio density functional theory (DFT) and post Hartree-Fock configuration interaction (CI) theoretical calculations. Comparisons of the results with other relevant isoelectronic atomic and molecular species are also presented in this section.

\section{\label{Exp+results}Experimental details and presentation of results}
\subsection{\label{Exp}Experimental procedures and details}
A detailed description of the merged-beam MAIA (Multi-Analysis Ion Apparatus) apparatus as well as the general experimental procedures used in the present work are given in Ref.~\citep{Bizau_2016} and our more recent works, e.g.~\citep{Mosnier_2025,Kennedy_2018}. Some specific details pertaining to the present work are given below and in Table~\ref{Expar}. The \ce{HS^{+}} or \ce{HCl^{+}} sample (parent) molecular ions were produced from \ce{H_2S} or \ce{HCl} gas, respectively, leaked into a permanent magnet electron cyclotron resonance ion source (ECRIS). The latter was excited by a 12.36 GHz microwave power supply and operated at the powers indicated in Table~\ref{Expar}. For each of the photoionisation experiments, the individual beam of the \ce{HS^+} or \ce{HCl^{+}} parent ion was extracted at $-4$ kV producing a velocity $v_{\text{4kV}} \ \text{ms}^{-1}$. A magnetic filter and electrostatic deflector, tuned to $v_{\text{4kV}}$ and a given nuclear mass, then selected and guided the sample ion (\ce{^1H^{32}_{16}S^+} or \ce{^1H^{35}_{17}Cl^+}) beam to the interaction region. The latter is a 0.57 m in length and 0.06 m in diameter cylinder subtending a solid angle of $8.7\times10^{-3}$ sr and an exit angle of 6.0$\degree$ at the (ion) entrance apex. The background pressure was $\simeq 2\times10^{-9}$ mbar for all the experiments. In the interaction region the parent ion beam was merged with the counter-propagating, left-handed circularly polarised, synchrotron radiation (SR) beam produced by an undulator and monochromatised by a 600 l/mm grating. The effects of the superposition of the short wavelength synchrotron radiation reflected in the second and third diffraction orders of the grating were evaluated and found to have negligible contributions on the final cross section results. In order to capture the greatest possible number of tagged ions (see below), an electrostatic (Einzel) lens of matched aperture follows the polarised cylinder. 

After the interaction region, the \ce{S^2+}or \ce{S^3+}, and \ce{Cl^2+} or \ce{Cl^3+}, photo-ion signals were counted with a microchannel plate following mass/charge analysis by a combination of a bending electro-magnet and an electrostatic deflector.  A high-voltage (tag) bias of -1 kV was applied to the interaction region in order to distinguish the photo-ions produced within the interaction region from those produced outside the region. In consequence, the singly charged parent positive ion beam saw its kinetic energy (KE) of 4 keV increased to 5 keV as it penetrated the interaction volume. The value of the speed of the \ce{HS^+} and \ce{HCl^{+}} parent ions in the interaction region is thus set to $V_C^{\ce{H^{32}S^+}}=1.72\times10^5$ ms$^{-1}$ and $V_C^{\ce{H^{35}Cl^+}}=1.65\times10^5$ ms$^{-1}$, respectively. Calibration of the apparatus has provided the following relationships between the electro-magnet current $I_A$, the resulting magnetic field strength $B_G$ and the kinetic energy $E_K$ of the ion (mass $m$) under analysis after the interaction region:
\begin{equation}\label{Magnet2}
B_G=31.845I_A-20.618=\frac{5.806}{q^{\prime}}\sqrt{mE_K}
\end{equation}
where $q^{\prime}$ is the final charge state of the fragment. $B_G$, $I_A$, $m$ and $E_K$ are in units of gauss, amp, amu and eV, respectively.

By measuring the photon and ion beam parameters, their overlap volumes (the Form factor in Table~\ref{Expar}), using calibrated photon and ion detectors and subtracting noise, it was possible to obtain the measured cross sections on an absolute basis \citep{Bizau_2016}. Finally, photon energy scans of the photo-ions count rates were carried out to map out the energy dependence of the photoabsorption cross sections. The photon energies were calibrated using a gas cell and known argon reference lines \citep{Ren_2011} and could be determined to within an accuracy of 40 meV. The relative uncertainty in the measured absolute cross sections is generally within 15\% \citep{Bizau_2016}.
  
 \begin{table}
\small
 \caption{\label{Expar} \ Sample values of main experimental parameters for the production of  the \ce{S^{2+}}, \ce{S^{3+}} and  \ce{Cl^{2+}}, \ce{Cl^{3+}} ionic fragments  at the $2p \rightarrow 3\sigma_d$ resonance energies of 175.0 and 212.5 eV  in \ce{HS^{+}} and \ce{HCl^{+}}, respectively}
  
 \begin{tabular*}{0.48\textwidth}{@{\extracolsep{\fill}}lll}
 \hline
 Experimental parameter & \ce{HS^{+}} & \ce{HCl^{+}} \\
 \hline
    Photoion count (cps) \ce{S^{2+}} \ce{Cl^{2+}}   & 31,860/10 s & 6680/4 s  \\
    Background count (cps) \ce{S^{2+}} \ce{Cl^{2+}}& 1810/10 s & 1260/4 s\\ 
    Photoion count (cps) \ce{S^{3+}} \ce{Cl^{3+}}   & 2876/10 s & 1470/10 s \\
    Background count (cps) \ce{S^{3+}} \ce{Cl^{3+}}& 100/10 s & 247/10 s \\ 
    Photodiode current \ce{S^{2+}} \ce{Cl^{2+}} ($\mu$A) & 160 & 44 \\
    Photodiode current \ce{S^{3+}} \ce{Cl^{3+}} ($\mu$A) & 160 & 44 \\
    Ion current \ce{S^{2+}} \ce{Cl^{2+}} (nA) & 106 & 595 \\
    Ion current \ce{S^{3+}} \ce{Cl^{3+}} (nA) & 107 & 570 \\
    Ion beam velocity $v$ (ms$^{-1})$ & $1.72\times10^{5}$ & $1.65\times10^{5}$ \\
    Tag voltage (kV) &  -1.0   &   -1.0  \\
    ECRIS RF power ($\mu$W) &  63    &    71     \\
    Channel plates efficiency $\epsilon$ & 0.56 & 0.56 \\
    Form factor \ce{S^{2+}} \ce{Cl^{2+}} (m$^{-1}$)   &   42,700    &    30,000\\
    Form factor \ce{S^{3+}} \ce{Cl^{3+}} (m$^{-1}$)   &   43,500    &    34,000\\
 \hline
 \end{tabular*}
\end{table}

\subsection{\label{Results}Principle of the measurements and results}
Upon absorption of a photon creating an inner-shell vacancy, the parent molecular ion will emit one (or several) Auger electron(s) and, simultaneously or sequentially, dissociate into atomic fragments. The KE of the fragments will necessarily include a contribution from the kinetic energy release (KER) inherent in the series of relaxation processes leading to nuclear dissociation of the parent molecular ion, see e.g. Ref.~\citep{Nenner_1996}. $E_K$ in eqn~\ref{Magnet2} comprises the KER value for the decay/dissociation process at play. This KE change is also accompanied by a change in the direction of travel of the fragments, away from the initial direction of the parent ion beam. These processes are key to our measurements for they underpin our analyses of the ionic fragment yield data and subsequent estimation of photo-absorption cross sections. In the Appendix Section at the end of the paper, we provide background details on the dynamics of formation of the dissociation fragments using the basic equations that express conservation of momentum and energy. We, thus, calculate, as a function of KER, both the expected KE changes and dissociation angles of the fragments with respect to the corresponding original values of the parent molecular ion beam. Notably, we show there that these additional\footnotemark[5] \footnotetext[5]{This is meant in reference to similar measurements performed on \emph{atomic} ion beams.}, purely molecular, effects do not affect our \ce{X^{2+}} and \ce{X^{3+}} fragments yield measurements in the specified experimental conditions, bar the unlikely case of very large KER values of many tens of eV's.

\begin{figure}
\centering
  \includegraphics[width=8.2cm]{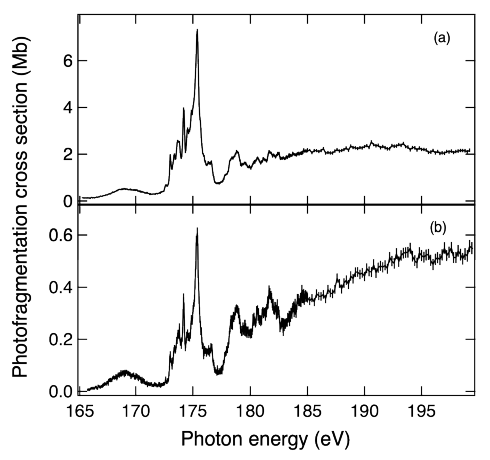}
  \caption{Cross sections for the \ce{HS^+} (sulfanylium) molecular ion for the photo-ion yield of (a) \ce{S^{2+}} ions and (b) \ce{S^{3+}} ions as a function of photon energy in the region of the sulfur 2p threshold .}
  \label{HS+}
\end{figure}

\begin{figure}
\centering
  \includegraphics[width=8.2cm]{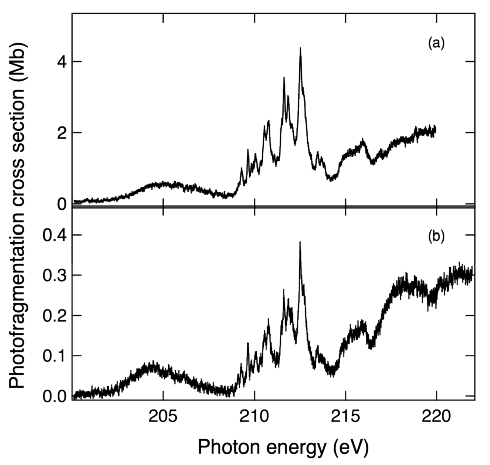}
  \caption{Cross sections of the \ce{HCl^+} (chloroniumyl) molecular ion for the photo-ion yield of (a) \ce{Cl^{2+}} ions and (b) \ce{Cl^{3+}} ions as a function of photon energy in the region of the chlorine 2p threshold .}
  \label{HCl+}
\end{figure}

More specifically in the present case, following the absorption of a photon of energy in the region of the \ce{X} $2p$ threshold, the \ce{HX^{+*}} parent molecular ion may either (1) directly dissociate and the resulting \ce{X^{+*}} ion single- or double-Auger decay to \ce{X^{2+}} or \ce{X^{3+}}, or (2) resonant/Auger decay with subsequent multiple dissociation routes of the \ce{HX^{+}^*}/\ce{HX^{2+}}$2p^{-1}$ mono/dication leading to \ce{X^{2+}}or \ce{X^{3+}} fragments\footnotemark[2] \footnotetext[2]{Although higher-order ionisation processes are possible, we limit our discussion to single and double ionisation.}or (3) decay radiatively (with possible subsequent unimolecular dissociation, e.g. \ce{HX^{+*} ->S +H^+ +}$h\nu^{\prime}$). Therefore, an appropriate summation of the measured cross sections for the total yield of the \ce{X^{+}}, \ce{X^{2+}}and \ce{X^{3+}} ions will provide a fair approximation of the photo-absorption cross section of the \ce{HX^+} molecular ion if pure radiative processes are ignored. We present these results for the \ce{X^{2+}} ions, \ce{X^{3+}} daughter ions\footnotemark[3] \footnotetext[3]{We have found no experimental evidence for the presence of \ce{HX^{2+}} or \ce{HX^{3+}} at the detection site.} of the \ce{HS^+} and \ce{HCl^+} parent ions in Figs~\ref{HS+} and ~\ref{HCl+}, respectively. These data represent the total contributions from the multiple core excitation/ionisation/dissociation routes (those leading to \ce{H^-} are not considered) as follows, where $\bar e$ is a free electron : 


\begin{equation}
\label{ionisationprocesses}
\ce{HX^{+} + $h\nu$ ->}\begin{cases}
                                  \textit{\ce{X^{+} + H^+ + $\bar e$}}\\
                                  \ce{X^{2+} + H + $\bar e$}\\
                                   \ce{X^{2+} + H^+ + $2\bar e$}\\
                                   \ce{X^{3+} + H +$2\bar e$}\\
                                   \ce{X^{3+} + H^+ + $3\bar e$}\\
\end{cases}
\end{equation}

\begin{figure}
\centering
  \includegraphics[width=8.2cm]{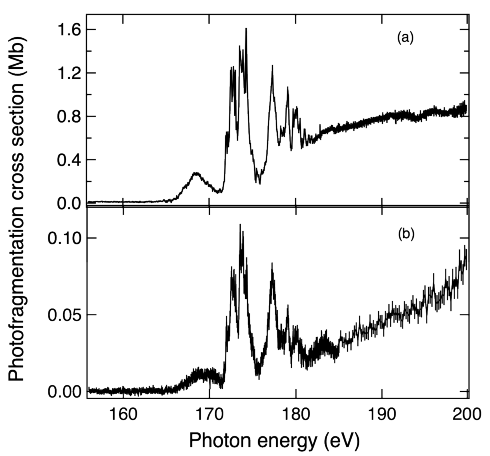}
  \caption{Cross sections of the \ce{H_2S^+} (sulfaniumyl) molecular ion for the photo-ion yield of (a) \ce{S^{2+}} ions and (b) \ce{S^{3+}} ions as a function of photon energy in the region of the sulfur 2p threshold .}
  \label{H2S+}
\end{figure}

Similar results are presented in Fig.~\ref{H2S+} for the triatomic \ce{H_2S^+} molecular ion. The \ce{X^{+}} photo-ion yield cannot be obtained in the present work as the 1 amu mass difference between the \ce{X^{+}} fragment and \ce{HX^{+}} parent ions is too small to be safely distinguished by our demerger/analysing apparatus. Therefore, the fragmentation routes ending on stable \ce{S^+} or \ce{Cl^+} atomic ions (italicised top line of eqn~\ref{ionisationprocesses}) are not amenable to experimental measurement. The total approximate photo-absorption cross sections, i.e. the sum of the single and double ionisation channels, are shown in Figs~\ref{ProtonationS+}(b),(c) and \ref{ProtonationCl+}(b) for \ce{HS^+}, \ce{H_2S^+} and \ce{HCl^+}, respectively.

In other experiments, carried out independently, we have measured the photoionisation cross sections of the \ce{S^+}~\citep{Mosnier_2025} and \ce{Cl^+}~\citep{Mosnier_2020} atomic ions. If it is assumed that the $2p$ photoionisation cross section for the \ce{HS^+} and \ce{HCl^+} molecular hydrides has a strong atomic character at photon energies markedly above threshold, then we may fruitfully compare the two sets of atomic and molecular data. At 190 eV photon energy, the \ce{S^+} continuum cross section is 3.4 Mb~\citep{Mosnier_2025} which is 18\% larger than the value obtained from~\ref{ProtonationS+}(b). In the case of the chlorine species, at the photon energy of 260 eV~\citep{Mosnier_2020}, this value is 15\%. As these differences are well within the combined error bars of the atomic and molecular measurements, they cannot be uniquely attributed to the contributions to the total molecular cross section of other (than \ce{X^2+} and \ce{X^3+}) ionisation/fragmentation channels.  For the triatomic \ce{H_2S^+} species, using the same normalisation as for \ce{HS^+}, the \ce{S^{2+}} and \ce{S^{3+}} cross sections are only 23\% and 1\% of the atomic \ce{S+} values, leaving a potential contribution of 76\% for the other ionisation/fragmentation channels. This seems compatible with the new physical conditions appearing in the case of the photoionisation of a triatomic molecule, viz. the number of possible fragmentation/ionisation routes following electronic relaxation is significantly increased leading to a larger selection of charge/mass ratios for the fragments.

In Sections~\ref{theor} and~\ref{Results}, the experimental data for the \ce{HS^+} and \ce{HCl^+} diatomics are analysed in the light of ab initio density functional theory (DFT) and post Hartree-Fock multiconfiguration self-consistent field theoretical calculations of the absorption oscillator strengths in the region of the $2p$ threshold. As similar calculations are not available for the \ce{H_2S^+} triatomic molecule,  the experimental data are discussed qualitatively without the support of theoretical simulations.

\begin{figure}
\centering
  \includegraphics[width=8.2cm]{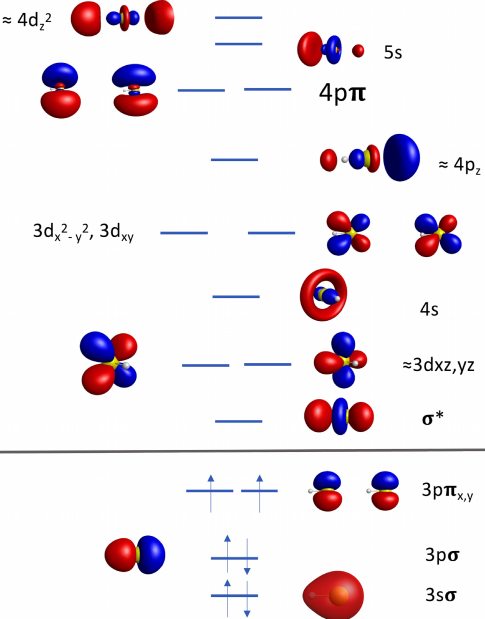}
  \caption{MO shapes and energy ordering \ce{HS^+}}
  \label{MOHS+}
\end{figure}

\begin{figure}
\centering
  \includegraphics[width=8.2cm]{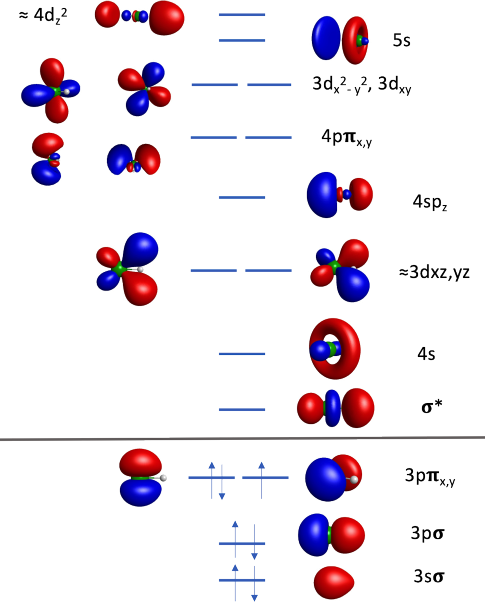}
  \caption{MO shapes and energy ordering \ce{HCl^+}}
  \label{MOHCl+}
\end{figure}

\section{\label{theor}Theoretical and computational aspects}
\subsection{\label{Elec}Calculation of electronic energies.}

The open-shell triplet HS$^{+}$ and doublet HCl$^{+}$ molecules belong to the C$_{{\infty} v}$ Point Group. Their valence-shell configurations are 3s$\sigma^{2}$3p$\sigma^{2}$3p$\pi^{2}$ and 3s$\sigma^{2}$3p$\sigma^{2}$3p$\pi^{3}$, respectively. All the calculations in this work were performed with the GAMESS(US) computer package \cite{GAMESS93}. Initial geometries were optimised using density functional theory (DFT) at the B3LYP level, which combines the Becke three-parameter hybrid exchange functional \cite{becke3} with the Lee-Yang-Parr gradient-corrected correlation functional \cite{lyp} including third-order Douglas-Kroll\cite{DK3} scalar relativistic corrections. The augmented correlation-consistent polarised core-valence quintuple-zeta (aug-cc-pCV5Z) basis set \cite{Woon1994,Pritchard2019a} was employed, with additional diffuse (4s, 4p, 4d) functions added for sulfur and chlorine atoms. The corresponding shapes and energy ordering of the topmost occupied and lowest unoccupied valence molecular (MO) orbitals are shown in Figs.~\ref{MOHS+} and ~\ref{MOHCl+}, for \ce{HS^+} and \ce{HCl+}, respectively.

The structural and spectroscopic properties of \ce{HS^+} and \ce{HCl^+} differ significantly from those of their \ce{HS} and \ce{HCl} neutral parents. Upon ionisation, both species exhibit slightly elongated bond lengths: 1.3694 {\AA}  for \ce{HS^+} and 1.3233 {\AA} for \ce{HCl^+}, compared to 1.34-1.36 {\AA} for \ce{HS} and 1.2746 {\AA} for \ce{HCl}. This elongation is indicative of a reduced electron density in the bonding region, consistent with the removal of an electron from a bonding or weakly antibonding orbital. Dipole moments increase upon ionisation, from 0.974 D (\ce{HS}) to 1.2531 D (\ce{HS^+}), and from 1.08 D (\ce{HCl}) to 1.2479 D (\ce{HCl^+}), reflecting the enhanced polarisation and charge separation in both the cationic states.
The fundamental vibrational frequencies are red-shifted relative to the neutral species: from approximately 2600 cm$^{-1}$ (\ce{HS}) to 2517.66 cm$^{-1}$ (\ce{HS^+}), and from 2940 cm$^{-1}$ (\ce{HCl}) to 2614 cm$^{-1}$ (\ce{HCl^+}). This decrease corroborates the weakening of the bond strength upon removal of one valence electron. When comparing the two cations, \ce{HCl^+} exhibits a shorter bond length and a larger vibrational frequency than \ce{HS^+}. This is consistent with the stronger \ce{H-Cl^+} bond and is likely driven by the higher electronegativity of chlorine relative to sulfur.



\subsection{\label{Abs}Simulation of absorption cross section spectra.}

The final-state energies and absorption intensities in the regions of the sulphur and chlorine (2p) L-edge were calculated in the equilibrium geometry of their respective ground state. 
The intensities (oscillator strengths) calculations were performed taking into account spin-orbit coupling (SOC) effects at the Breit-Pauli level of theory and using a post-Hartree-Fock configuration interaction approach restricted to single-excitations (CI-S), as implemented in the GAMESS code. These effects are intrinsic to the sulfur and chlorine 2p core levels, where spin-orbit splitting leads to energetically distinct 2p$_{3/2}$ and 2p$_{1/2}$ components, with differences for sulphur and chlorine of $\simeq$1.2 and 1.6 eV, respectively. The inclusion of such coupling is essential for accurately modelling the fine structure observed in core-level spectra.
The spin-orbit coupling calculations were performed on variational CI wavefunctions, using optimized Hartree-Fock [\ce{HS^{2+}}]/[\ce{HCl^{2+}}]2s$^{-1}$ molecular orbitals (MOs) as initial guess orbitals for the SOC-CI calculations. The core-excited reference \ce{S}/\ce{Cl} 2s$^{-1}$ configurations were chosen to avoid the artificial preferential orientation effects that can arise in spin-orbit computations for \ce{S}(2p)/\ce{Cl}(2p)$^{-1}_{x,y,z}$. Given that the \ce{S}(2s)/\ce{S}(2p) and \ce{Cl}(2s)/\ce{Cl}(2p) orbitals are each (independently) nearly close in energy, equivalent relaxation effects are expected. 

To account for the nuclear motion caused by core-level excitation of the low-lying excited states of interest, numerical gradients were calculated for the core-excited states in the ground state geometries. The intensities of the vibrational progressions were calculated within the Franck-Condon approximation.

\section{\label{Results}Analyses and discussions}
\subsection{\label{exp-theoHS+}\ce{HS^+}photo-absorption cross sections}
\subsubsection{\label{Compexp-theoHS+}Comparison theoretical-experimental data}
\begin{figure*}
\centering
  \includegraphics[width=12cm]{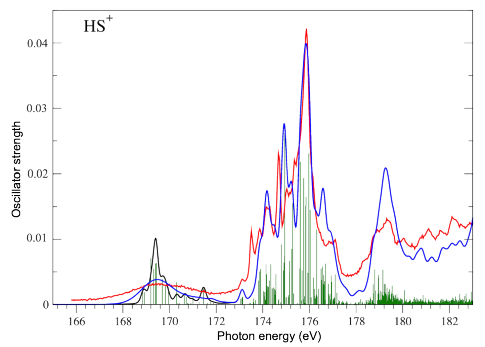}
  \caption{Comparison of experimental (red trace) and theoretical (blue trace) relative photoabsorption cross sections of \ce{HS^+} as a function of photon energy in the region of the $2p$ threshold. The green-coloured vertical sticks are the calculated oscillator strengths. The resonances between 168 and 172 eV (black and blue traces) are discussed in the main text of the paper.}
  \label{ExpTheoHS+}
\end{figure*}

\begin{figure}
\centering
  \includegraphics[width=8.2cm]{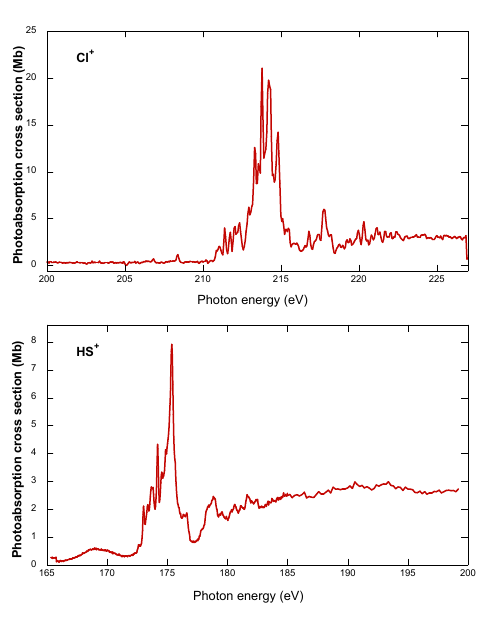}
  \caption{Total photo-absorption cross section in the region of the 2p sub-shell ionisation threshold for the isoelectronic \ce{Cl^+} atomic ~\citep{Mosnier_2020} and  \ce{HS^+} molecular ions (top and bottom, respectively)}
  \label{ComparisonCl+vsHS+}
\end{figure}

To enable a direct comparison between the theoretical calculations and the experimental data, a synthetic spectrum was generated by assigning normalised Gaussian profiles to the calculated oscillator strengths. The full width at half maximum (FWHM) of the Gaussians approximated the experimental spectral band pass (BP) of 130 meV in the photon energy region near the $2p$ \ce{HS^+} ionisation threshold. The resulting photoabsorption relative cross section spectrum is shown in Fig.~\ref{ExpTheoHS+} superimposed onto the corresponding experimental trace and the calculated oscillator strengths shown as vertical sticks. It is seen that the agreement between theory and experiment is very satisfactory, both in terms of the energy positions and relative intensities of the resonance features. The experimental spectrum is displayed on an absolute cross section scale (Mb) in the lower and middle panels of Figs.~\ref{ComparisonCl+vsHS+} and~\ref{ProtonationS+}, respectively.

For \ce{HS^+}, the low-energy region of the theoretical absorption spectrum involves final states of the type $2p^5\pi^3$, see Table~\ref{TableHS+}. These states exhibit very weak transition intensities and, thus, should be barely visible in the experimental spectrum (this spectral region was actually not scanned in this work). The first experimental spectral feature of noticeable intensity in Fig.~\ref{ExpTheoHS+}appears around 169 eV and consists of a broad, featureless band. This band is attributed to configurations of the form $2p^5\pi^2\sigma^{*1}$, extending over approximately 4 eV (black curve). Its large width reflects the dissociative character of the excited states which involves breaking of the \ce{HS^+} bond. This interpretation is supported by the calculated energy gradient of $\sim$ 7 eV/{\AA}, in the equilibrium geometry of the ground state, confirming the strongly dissociative nature of the $2p^5\pi^2\sigma^{*1}$ configuration. When the bond-breaking effect in the Franck-Condon region is accounted for, the broadened theoretical profile (blue curve) closely matches the experimental shape.
Between 173 and 177 eV, the most intense part of the spectrum features a series of well-resolved peaks corresponding to $2p \rightarrow 3d$ transitions, leading to final states of configuration $2p^53p^23d^1$.
At around 178 eV, the spectrum enters the region of $2p \rightarrow 4d$ transitions, and rapidly converges toward the $2p_{1/2,3/2} \rightarrow \epsilon d$ ionisation continua, with a theoretical threshold energy of  $\approx$ 183 eV for HS$^{+}_{2p_{3/2}}$. Resonance assignments below the 2p thresholds are summarised in Table~\ref{TableHS+} which gives more details on the electronic configurations of the final states.

\begin{table}
\small
  \caption{\ Spectral assignment of the main resonance features in the photoabsorption spectrum of \ce{HS^+} near the $2p$ threshold photon energy region}
  \label{TableHS+}
  \begin{tabular*}{0.48\textwidth}{@{\extracolsep{\fill}}ll}
    \hline
    Photon energy (eV) & Main electronic configuration  \\
    \hline
    161.0-162.0 & $2p^5\pi^2\pi^1$  \\
    168.0-172.0 & $2p^5\pi^1\pi^1\sigma^{*1}$  \\
    173.6-174.5 & $2p^5\pi^1\pi^1[4s/3d_{xz,yz}]^1$ \\
    174.6-175.6 & $2p^5\pi^1\pi^1[3d_{xz,yz}]^1$ \\
    175.6-176.4 & $2p^5\pi^1\pi^1[4p_z/3d_{xy},3d_{x^2-y^2}]^1$ \\
    176.4-177.4 & $2p^5\pi^1\pi^1[4p_z,4p_{\pi}/3d_{xy},3d_{x^2-y^2}]^1$ \\
    178.8-180.0 & $2p^5\pi^1\pi^1[4d,5s,5p]^1$ \\
    \hline
  \end{tabular*}
\end{table}

\subsubsection{\label{SIDIHS+}Ionic and fragmentation decay channels}
The total \ce{HS^+} photoabsorption cross section in the region of the 2p threshold essentially comprises the contributions from all the decay channels, cf eqn~\ref{ionisationprocesses}, terminating on the \ce{S^2+} and \ce{S^3+} ions in electronic states stable against further ionisation, cf Figs.~\ref{HS+}(a) and (b), resp. As our experimental technique would need to be completed by electron spectrometry to gain insight into the dynamics of nuclear dissociation vs electronic decay, we cannot give definitive conclusions in this regard. We have, however, shown in the case of the \ce{S^+} atomic ion that the largest autoionisation rate of the $2p^53s^23p^33d$ configuration is via the $2p^53s^23p^33d\rightarrow2p^63s^23p3d+\varepsilon \bar e$ spectator Auger transition ($\varepsilon \bar e$ represents the continuum electron)~\citep{Mosnier_2025}. This may suggest that \ce{HX^{+} + $h\nu$ ->}\ce{X^{2+} + H + $\bar e$} is a dominant decay channel for the resonant Auger processes contributing to Fig.~\ref{HS+}(a) as the $3p$ valence electrons involved in the Auger process are localised on the sulphur atom. A significant proportion of broad, continuum-like processes (as opposed to the sharp, atomic-like, $3d_\sigma$ resonances) is also seen to contribute to the cross section in the single-ionisation channel above the photon energy of $\approx$ 180 eV, i.e close to the theoretical threshold of $\approx$ 183 eV for HS$^{+}_{2p_{3/2}}$. This may indicate different decay channels for the $nd_\sigma$ and $(n+1)s_\sigma,\, n>3$ Rydberg states with comparatively shorter lifetimes or that they possibly are located in the autoionisation continuum. Regarding the behaviour of the cross sections due to the double ionisation channels shown in Fig.~\ref{HS+}(b), we note the almost replicated spectral distributions in Figs.~\ref{HS+}(a) and (b) below the 2p thresholds, while the continuum part features a comparatively sharper increase in Fig.~\ref{HS+}(b). The latter observation is straightforwardly attributable to the onset of the direct Auger process above the 2p threshold. In the resonance region, we may generally invoke direct double Auger ($L_{2,3}-MMM$ here), cascade Auger or initial state correlations (shake-off) processes to account for the double-ionisations. Extensive multi coincidence techniques would be needed to unravel the decay dynamics from the experimental data, see e.g. Ref.~\citep{Journel_2008} and indicate whether dissociation followed by autoionisation of the fragments or molecular Auger decay is the dominant process. In the present case, since the 2p vacancy is necessarily filled first as part of any Auger process, a limited number of M-shell emission schemes would be available for the second Auger process as part of the double Auger cascade.    

\subsubsection{\label{IsoelectronicHS+}Isoelectronic comparison \ce{Cl^+}-\ce{HS^+}}
Figure~\ref{ComparisonCl+vsHS+} shows the \ce{HS^+} photoabsorption cross section spectrum against that of the isoelectronic \ce{Cl^+}~\citep{Mosnier_2020} (united atom) species in the region of their respective 2p thresholds. As expected the spectral distributions of the photoabsorption cross sections are similar in both spectra with a predominance of the oscillator strength into the $2p \rightarrow 3d$ resonances compared to the valence and Rydberg states. The broad $2p \rightarrow \sigma^*$ resonance uniquely features in the molecular spectrum.

\begin{figure}
\centering
  \includegraphics[width=8.2cm]{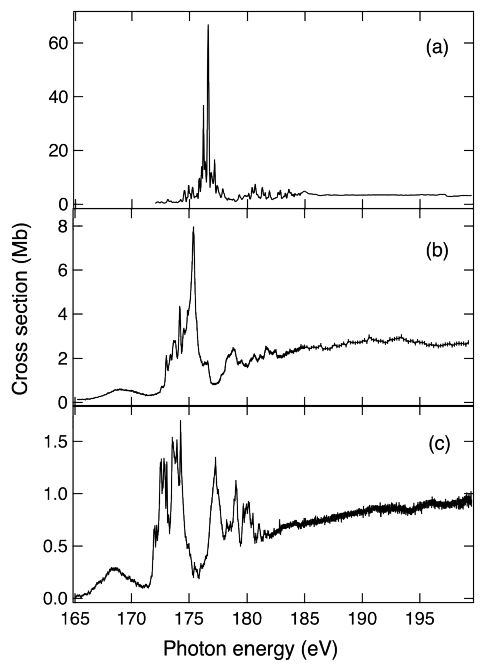}
  \caption{Total photo-absorption cross section in the region of the 2p sub-shell ionisation threshold for the \ce{S^+} atomic (a), \ce{HS^+} (b) and \ce{H_2S^+} (c) molecular ions.}
  \label{ProtonationS+}
\end{figure}

\subsubsection{\label{ProtonationS+HS+}Effects of protonation on \ce{S^+}}
Figure~\ref{ProtonationS+} compares the near 2p threshold photoabsorption spectra of \ce{S^+} (ref.~\citep{Mosnier_2025}), \ce{HS^+} and \ce{H_2S^+}. The figure shows the effects of the molecular field resulting from the formation of a chemical bond with one hydrogen atom (\ce{HS^+}) or two hydrogen atoms (\ce{H_2S^+}), on the atomic potential of the \ce{S^+} ion. The enhancement of the absorption strength into the valence and Rydberg states, relative to that of the $2p \rightarrow 3d_{\sigma}$ resonance, is very noticeable in \ce{H_2S^+}. Most of the resonance strength of the $2p\rightarrow nd$ series is concentrated in the $2p\rightarrow 3d$ excitations in \ce{S^+} (re.~\citep{Mosnier_2025}), indicative of the centrifugal barrier effects of the 3d orbital. The $2p \rightarrow 3d_{\sigma}$ resonances in \ce{HS^+} and \ce{H_2S^+} are seen to be sensitive markers of the modified chemical/electronic environment. Additionally, we remark that the high-resolution $L_{2,3}$ photoabsorption data of \ce{H_2S} shown in Ref~\cite{Hudson_1994} closely resembles the total photoion yield spectrum of Fig.~\ref{ProtonationS+}(c) for \ce{H_2S^+}. We note that the main decay channels of neutral \ce{H_2S} following $2p\rightarrow \sigma^*$ resonant excitation is via neutral dissociation into \ce{H + HS^*} followed by molecular Auger decay of \ce{HS^*}~\citep{H_Aksela_1992a,H_Aksela_1992c,Le-Guen_2007}. Conversely, the sulfur 2p photoabsorption spectrum of \ce{SF_6}~\citep{Hudson_1993,Stener_2011} differs markedly from \ce{H_2S} or \ce{H_2S^+}, mirroring the effects of the fluorine cage.

\subsection{\label{exp-theoHCl+}\ce{HCl^+}photo-absorption cross sections}
In the following, we shall adopt a pattern similar to Sec.~\ref{exp-theoHS+} for the order of the discussion of the main features of the \ce{HCl+} data. Some of the details will therefore not be repeated again in this section and the reader will easily make reference to Sec.~\ref{exp-theoHS+} for cross-comparisons. 

\subsubsection{\label{Compexp-theoHCl+}Comparison theoretical-experimental data}
\begin{figure*}
\centering
  \includegraphics[width=12.0cm]{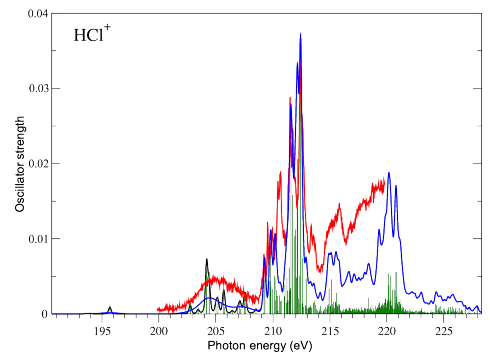}
  \caption{Comparison of experimental (red trace) and theoretical (blue trace) relative photo-absorption cross sections of \ce{HCl^+} as a function of photon energy in the region of the $2p$ threshold. The green-coloured vertical sticks are the calculated oscillator strengths. The resonances between 200 and 210 eV (black and blue traces) are discussed in the main text of the paper.}
  \label{HCl+ExpTheo}
\end{figure*}

\begin{figure}
\centering
  \includegraphics[width=8.2cm]{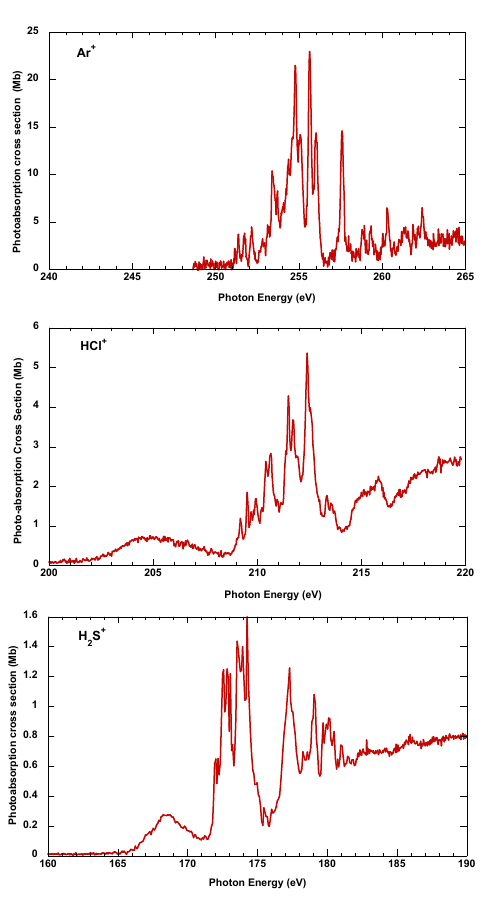}
  \caption{Total photo-absorption cross section in the region of the 2p sub-shell ionisation threshold for the isoelectronic \ce{Ar^+} atomic~\cite{Blancard_2012}  and \ce{HCl^+} and \ce{H_2S^+} (from top to bottom) molecular ions.}
  \label{ComparisonAr+HCl+H2S+}
\end{figure}

The comparison between the theoretical calculations and the experimental data is shown in the synthetic spectrum of Fig.~\ref{HCl+ExpTheo}. The full width at half maximum (FWHM) of the Gaussians was set equal to 150 meV over the 200-230 eV range. The actual experimental BP was 165 meV at 206 eV photon energy. The resulting photoabsorption relative cross section simulation is again superimposed onto the corresponding experimental trace and the calculated oscillator strengths. It is seen again that the agreement between theory and experiment is very satisfactory, both in terms of the energy positions and relative intensities of the resonance features. The middle and lower panels of Figs.~\ref{ComparisonAr+HCl+H2S+} and~\ref{ProtonationCl+} show the absolute experimental cross sections.

The low-energy region near 195 eV corresponds to final states of type $2p^5\pi^4$. These were not observed in the experiments. This is followed by a broad, structureless feature centred around 205 eV, attributed to configurations of the type $2p^5\pi^3\sigma^{*1}$. The broadening again reflects a dissociative excited state, involving the breaking of the H-Cl bond. The energy gradient computed at the equilibrium geometry of the \ce{HCl^+} ground state is calculated to be $\sim\, -$ 9 eV/\AA, confirming the strongly dissociative nature of the $2p^5\pi^3\sigma^{*1}$ configuration. Here again, the inclusion of dissociation-induced broadening in the computer simulation (blue curve between 200 eV and 210 eV) results in an excellent agreement with the experimental spectrum. In the 206-214 eV region, the most intense part of the spectrum reveals a series of sharp peaks associated with $2p \rightarrow 3d_{\sigma}$ transitions, leading to atomic-like final states of the form $2p^53s^23p^33d^1$.
Above 214 eV, the spectrum progresses into the $2p \rightarrow$ Rydberg transition region featuring only a broad and highly blended resonance structure.  The HCl$^{+}_{2p_{3/2}}$ ionisation threshold is theoretically established at $\approx$ 220 eV. Resonance assignments are summarised in Table~\ref{TableHCl+} with details on the electronic configurations of the final states.

\begin{table}
\small
  \caption{\ Spectral assignment of the main resonance features in the photo-absorption spectrum of \ce{HCl^+} near the $2p$ threshold photon energy region}
  \label{TableHCl+}
  \begin{tabular*}{0.48\textwidth}{@{\extracolsep{\fill}}ll}
    \hline
    Photon energy (eV) & Main electronic configuration  \\
    \hline
    191.0-196.0 & $2p^5\pi^4$  \\
    200.0-208.0 & $2p^5\pi^3\sigma^{*1}$  \\
    209.0-211.0 & $2p^5\pi^3[4s/3d_{xz,yz}/3d_{x^2-y^2}]^1$ \\
    210.0-215.0 & $2p^5\pi^3[3d_{xz,yz}/3d_{x^2-y^2}/4p_z/4p_{\pi^*}]^1$ \\
    215.0-216.0 & $2p^5\pi^3[4d,5s,5p]^1$ \\
    220.0 & Admixture of Rydberg states \\
    225.0 & Admixture of Rydberg states \\
    \hline
  \end{tabular*}
\end{table}

\subsubsection{\label{SIDIHCl+}Ionic and fragmentation decay channels}
The total \ce{HCl^+} photoabsorption cross section in the region of the 2p threshold essentially comprises the contributions from all the decay channels, cf eqn~\ref{ionisationprocesses}, terminating on the \ce{Cl^2+} and \ce{Cl^3+} ions in electronic states stable against further ionisation, cf Figs.~\ref{HCl+}(a) and (b), resp. The overall behaviour is quite similar to \ce{HS+}. That is to say, (1) an intense group of sharp atomic-like $3d_\sigma$ resonances followed to higher photon energies by broad, smeared resonance patterns corresponding to the valence/Rydberg states and (2) almost identical spectral patterns in the single and double ionisation spectra bar a direct Auger enhancement above the $2p$ threshold (218-220 eV) in the double ionisation channel. All the comments and suggestions of section~\ref{SIDIHS+} can be transposed and repeated for the case of \ce{HCl^+}, with the exception of the comments based on the largest autoionisation rate of the $2p^53d$ core-excited configuration since we do not have similar calculations for \ce{Cl^+}.

\subsubsection{\label{IsoelectronicAr+HCl+}Isoelectronic comparison \ce{Ar^+}, \ce{HCl^+}, \ce{H_2S^+}}
Figure~\ref{ComparisonAr+HCl+H2S+} shows the \ce{HCl^+} photoabsorption cross section spectrum against that of the isoelectronic \ce{Ar^+}~\citep{Blancard_2012} (united atom) species in the region of their respective 2p thresholds. The \ce{Ar^+} spectrum was obtained with a 170 meV photon energy BP. All three spectra were obtained with different BP's and, thus, resonance widths or strengths cannot be directly compared. As expected the spectral distributions of photoabsorption cross sections are quite similar in the \ce{Ar^+} and \ce{HCl^+} spectra with a predominance of the discrete oscillator strength into the $2p \rightarrow 3d$ resonances compared to the valence and Rydberg states. The broad $2p \rightarrow \sigma^*$ resonance uniquely features in both the molecular spectra. This is somewhat at variance with the $2p$ photoabsorption cross section profile for \ce{H_2S^+}  that shows a marked enhancement of the absorption strength into the valence and Rydberg states (see section ~\ref{ProtonationS+HS+}).

\begin{figure}
\centering
  \includegraphics[width=8.2cm]{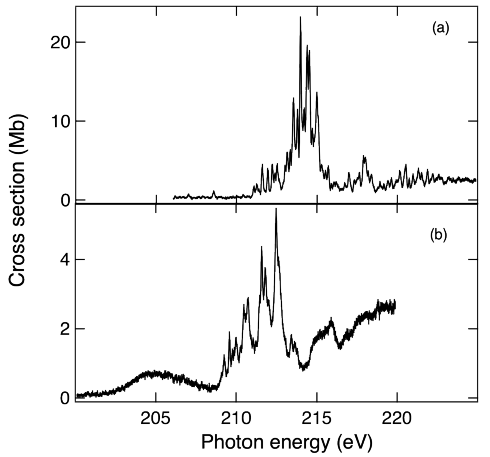}
  \caption{Total photo-absorption cross section in the region of the 2p sub-shell ionisation threshold for the \ce{Cl^+} atomic (a) and \ce{HCl^+} (b) molecular ions. The figure shows the effect of single protonation on the atomic potential of the \ce{Cl^+} ion.}
  \label{ProtonationCl+}
\end{figure}

\subsubsection{\label{ProtonationCl+HCl+}Effects of protonation on \ce{Cl^+}}
 Figure~\ref{ProtonationCl+} shows the effects of the molecular field resulting from the formation of a chemical bond with one hydrogen atom (\ce{HS^+}) on the atomic potential of the \ce{Cl^+} ion. Here again, the $2p \rightarrow nd_{\sigma}$ resonances in \ce{HCl^+} are seen to be sensitive markers of the modified chemical/electronic environment. We note that the total photoion yield spectrum of Fig.~\ref{ProtonationCl+}(b) for \ce{HCl^+} exhibits similarities with the total ion yield spectra of \ce{HCl} obtained using time-of-flight mass spectrometry~\citep{H_Aksela_1992b} or electron spectroscopy~\citep{Kivimaki_1993}. The \ce{HCl} absolute photoabsorption cross sections~\citep{Ninomiya_1981} of a few megabarns below and rising to the 2p edges are compatible with the measurements of figure~\ref{ProtonationCl+}(b). Additionally, it is of interest to remark that in \ce{HCl}, the $2p\rightarrow \sigma^*$ resonance decays via fast neutral dissociation followed by atomic \ce{Cl} resonance Auger decay, whereas the main decay channel for the Rydberg states following $2p$ excitation is via the molecular resonance Auger process with shake-up contributions~\citep{Kivimaki_1993}.





\section{\label{conclusions}Conclusions}
Monochromatised SOLEIL synchrotron radiation ($h\nu$) was used in a photon-ion merged beam apparatus to measure the absolute cross sections of the \ce{HS^{+} + $h\nu$ -> S^{2+}},  \ce{HS^{+} + $h\nu$ -> S^{3+}},  \ce{HCl^{+} + $h\nu$ -> Cl^{2+}}, \ce{ HCl^{+} + $h\nu$ -> Cl^{3+}}, \ce{ H_2S^{+} + $h\nu$ -> S^{2+}} and \ce{ H_2S^{+} + $h\nu$ -> S^{3+}} ionic photofragmentation reactions over the photon energy regions spanning the energies of the $2p$ ionisation thresholds of sulphur ($\sim$ 180 eV) and chlorine ($\sim$ 220 eV). The absorption oscillator strengths ($f$-values) of the \ce{HS^+} and \ce{HCl^+} $2p$ core excitations to valence and Rydberg states were computed using extensive ab initio density functional theory (DFT) and post-Hartree-Fock configuration interaction calculations including spin-orbit coupling. Very good agreements were obtained between the \ce{HS^{+}} and \ce{HCl^{+}} synthetic photoabsorption cross section spectra generated from the $f$-values and the summed (\ce{S^{2+}}+\ce{S^{3+}}) and (\ce{Cl^{2+}}+\ce{Cl^{3+}}) photofragmentation cross section spectra, respectively. Detailed comparison between the theoretical and experimental results enabled the electronic configuration assignment of some of the core-hole states, notably in the $2p^{-1}3d_\sigma$ spectral regions that contain several sharp resonances. Comparison of photoionisation data along the isoelectronic series starting at the united atom, ie the \ce{Ar^+}, \ce{HCl^+}, \ce{H_2S^+} and \ce{Cl^+}, \ce{HS^+} series, as well the effects of protonation along the \ce{S^+}, \ce{HS+}, \ce{H_2S^+} and \ce{Cl^+}, \ce{HCl^+} series, were considered. Common features in all the molecular spectra along these series included (1) an intense and broad $2p\longrightarrow \sigma^{\star}$ resonance, indicative of strongly dissociating upper states and (2) a marked atomic character for the $2p\longrightarrow 3d_{\sigma}$ resonances.


\section*{Data availability}
The primary data that support all the findings of this study are available in the published article. The manuscript is deposited in DORAS, the DCU Research Repository, at the following URL: \url{https://doras.dcu.ie/xxxxx/}. Additional data are available from the corresponding author upon any reasonable request

\section*{Conflicts of interest}
There are no conflicts to declare.

\section*{\label{Appen}Appendix}

The absorption of a single photon of energy $E$ in an inner electron subshell will excite the parent molecule (\ce{HX^+} in this work) to an autoionising resonance level. The subsequent decay of the latter by electron emission (Auger process) will usually leave the doubly-ionised molecule in a dissociative final state with excess internal energy $\epsilon_\text{int}$ above the asymptotic final potential energy of the fragments $\epsilon_\text{int}^{\ce{H}}$ and $\epsilon_\text{int}^{\ce{X}}$ ( \ce{HX -> H + X} fragmentation is assumed). This excess energy is retrieved in the form of kinetic energy imparted to the dissociation fragments. The sum of the KE for all the fragments constitutes the kinetic energy release (KER), i.e $\epsilon_\text{int}-(\epsilon_\text{int}^{\ce{H}}+\epsilon_\text{int}^{\ce{X}})=E_{\text{KER}}$. If the fragments are all electrically charged, the final step of this process -the release of the fragments into the continuum of kinetic energies- has historically been interpreted as a Coulombic explosion \citep{Carlson_1966}. See eqn~\ref{ionisationprocesses} for the ionisation channels where Coulomb explosion is possible in the present case. The analyses presented below are valid for neutral or charged fragments.

\begin{figure}
\centering
  \includegraphics[width=8.2cm]{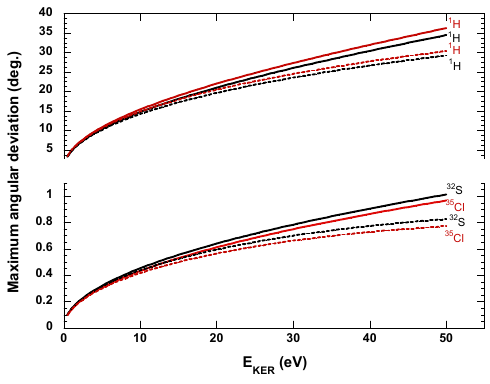}
  \caption{Maximum deviation angles from the direction of the parent beam for the \ce{^{32}S}, \ce{^{35}Cl} and \ce{^{1}H} dissociation fragments (solid lines) and corresponding deviation angles for the other daughter (broken lines) as a function of $E_{\text{KER}}$ (eV). Colour code: black and red traces for the \ce{^1H^{32}S^{+}} and \ce{^1H^{35}Cl^{+}}, parent molecular ion, respectively. (See text for details).}
  \label{Scattering angles}
\end{figure}

The kinematics of the dissociation fragments is determined solely by the laws of conservation of momentum and energy and similar to the spontaneous disintegration of a heavy particle into fragments of smaller masses \citep{Landau_1976,Laskin_2001, Martins_2021}. In the centre of mass (CM) frame of reference, the fragments move away from each other with equal and opposite momenta $p_0$ such that $E_\text{KER}=p_0^2/2\mu$ where $\mu$ is the reduced mass of the two fragments $1/\mu=1/m_{\ce{H}}+1/m_{\ce{X}}$ and $p_0=m_{\ce{H}}v_0^{\ce{H}}=m_{\ce{X}}v_0^{\ce{X}}$ with $v_0^{\ce{H}}$ and $v_0^{\ce{X}}$ the speed of the \ce{H} and \ce{X} fragments in the CM system, respectively. If the velocity of the parent molecular ion, i.e. the sample beam, in the laboratory (lab) frame is $\vec{V}_C^{\ce{HX}}$ then the following vector equations hold (see Fig.14 of Ref. \citep{Landau_1976})
\begin{equation}\label{vectors}
\vec{v}^{\thinspace \ce{H}}-\vec{V}_C^{\ce{HX}}=\vec{v}_{0}^{\thinspace \ce{H}}\  \text{and}\  \vec{v}^{\thinspace \ce{X}}-\vec{V}_C^{\ce{HX}}=\vec{v}_{0}^{\thinspace \ce{X}},
\end{equation} 
where $\vec{v}^{\thinspace \ce{H}}$ and $\vec{v}^{\thinspace \ce{X}}$ are the velocities of the \ce{X} and \ce{H} fragments in the lab frame, respectively. Solving these equations for $v^{\ce{H}}$ or $v^{\ce{X}} \equiv{v}$ yields,
\begin{equation}\label{speed}
v=V_C\cos\theta\pm v_0\sqrt{1-\frac{V_C^2}{v_0^2}\sin^2\theta}, 
\end{equation}
where $V_C\equiv V_C^{\ce{HX}}$ and $\theta\equiv \theta^{\ce{H}}\ \text{or}\ \theta^{\ce{X}}$ is the angle at which the \ce{X} or \ce{H} dissociation fragment moves away from the direction $\vec{V_C}$ of the parent ion beam\footnotemark[6] \footnotetext[6]{More precisely, $\theta$ is half the aperture cone angle of the spherical cone of radius $v_0$ centred on the direction $\vec{V_C}$ of the parent ion beam.}. If $V_C>v_0$ (in the present case $V_C\gg v_0$), the fragment can only move forward at any angle $\theta$ such that $0\leq \theta \leq\theta_{\text{max}}$ (see Fig.14 of Ref. \citep{Landau_1976}) with $\sin\theta_{\text{max}}=v_0/V_C$ . 
The behaviour of $\theta_{\text{max}}$ for the fragments (daughters) produced following dissociation of the \ce{^1H^{32}_{16}S^+} and \ce{^1H^{35}_{17}Cl^+} parent ions is presented using solid lines in Fig.~\ref{Scattering angles} as a function of $E_{\text{KER}}$. The range of $E_{\text{KER}}$ values, from a fraction of an eV to several ten's of eV's, is typical for the inner-shell photoionisation processes in hydrides we are considering in the present work, e.g. Refs~\citep{Puglisi_2018,Martins_2021}. Both daughters cannot simultaneously dissociate at their respective $\theta_{\text{max}}$ value. If, for argument's sake, the \ce{X} fragment dissociates at $\theta_{\text{max}}^{\ce{X}}$ then the \ce{^1H} daughter will dissociate at an angle $\theta^{\ce{^1H}}<\theta_{\text{max}}^{\ce{^1H}}$. The value of $\theta^{\ce{^1H}}$ is then obtained from considerations of the geometry of the dissociation process in the CM frame~\citep{Landau_1976} as follows. The deviation angle $\theta^X_0$ of the heavy fragment in the CM frame corresponding to its maximum deviation angle in the lab frame $\theta^X_{\text{max}}$ is such that
\begin{equation}
 \cos\theta^{\ce{X}}_0 =-\sin\theta^{\ce{X}}_{\text{max}} \, \text{with} \ \theta^{\ce{^1H}}_0+\theta^{\ce{X}}_0=\pi
 \end{equation}
 The \ce{^1H} deviation angles in the lab frame can now be obtained from 
 \begin{equation}
 \tan\theta^{\ce{^1H}} =\frac{v_0^{\ce{^1H}}\sin\theta_0^{\ce{^1{H}}}}{V_C+v_0^{\ce{^1H}}\cos\theta_0^{\ce{^1{H}}}}.
 \end{equation}
We have calculated these angles and their values are represented as broken lines in Fig.~\ref{Scattering angles}. For a given pair of daughters at a given value of $E_{\text{KER}}$, the sum (solid line + broken line) of the angular values gives the maximum (recoil) angle between the directions of motion of the two fragments as they move away from one another. The speed of the \ce{H} atom in the lab frame can now be obtained by substituting the value of $\theta^H$ in equation~\ref{speed}.
We see from Fig.~\ref{Scattering angles} that, as expected from the significant mass difference between the fragments, $\theta_{\text{max}}^{\ce{X}}$ is only a fraction of a degree while it is several ten's of degrees for the \ce{H} fragments. From the value of the construction parameters of our apparatus (see Section~\ref{Exp+results}), it is evident that all the heavy fragments will be transmitted through to the mass analysis/detection stage, while this will not be the case for most of the hydrogen fragments.

\begin{figure}
\centering
  \includegraphics[width=8.2cm]{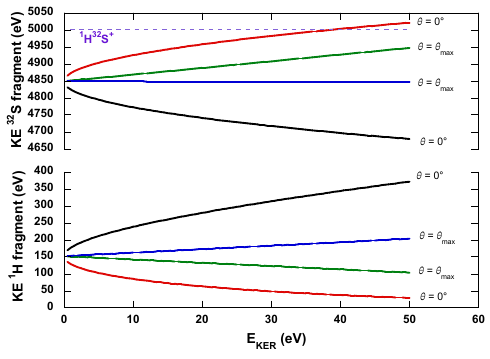}
  \caption{Kinetic energy (KE) of the \ce{^{32}S} (upper part) and \ce{^{1}H} (lower part) dissociation fragments as a function of $E_{\text{KER}}$ (eV) at $\theta=0\degree \ \text{and} \ \theta=\theta_{\text{max}}\degree$ dissociation angles. The purple coloured broken line at 5 keV represents the kinetic energy of the \ce{^1H^{32}S^{+}} parent molecular ion, ie. $V=1.709\times10^5$ ms$^{-1}$. Black solid lines: $KE_{\ce{^{32}S}}=\frac{1}{2}m_{\ce{^{32}S}}(V_C-v_0^{\ce{^{32}S}})^2$ and $KE_{\ce{^{1}H}}=\frac{1}{2}m_{\ce{^{1}H}}(V_C+v_0^{\ce{^{1}H}})^2$; red solid lines: $KE_{\ce{^{32}S}}=\frac{1}{2}m_{\ce{^{32}S}}(V_C+v_0^{\ce{^{32}S}})^2$ and $KE_{\ce{^{1}H}}=\frac{1}{2}m_{\ce{^{1}H}}(V_C-v_0^{\ce{^{1}H}})^2$, with $v_0^{\ce{^{32}S}}=427.2\sqrt{E_{\text{KER}}(\text{eV})}$ (ms$^{-1}$) and $v_0^{\ce{^{1}H}}=32v_0^{\ce{^{32}S}}$. Green solid lines: $KE_{\ce{^{1}H}}=\frac{1}{2}m_{\ce{^{1}H}}(V_C\cos \theta_{\text{max}}^{\ce{^{1}H}})^2$ with $\theta_{\text{max}}^{\ce{^{1}H}}$ shown as the \ce{^{1}H} black solid line on Fig.~\ref{Scattering angles}; see main text for calculation of $KE_{\ce{^{32}S}}$ in this case; blue solid lines: $KE_{\ce{^{32}S}}=\frac{1}{2}m_{\ce{^{32}S}}(V\cos \theta_{\text{max}}^{\ce{^{32}S}})^2$ with $\theta_{\text{max}}^{\ce{^{32}S}}$ shown as the \ce{^{32}S} black solid line on Fig.~\ref{Scattering angles}; see main text for calculation of $KE_{\ce{^{1}H}}$ in this case and additional background details.}
  \label{KEFragments1}
\end{figure}

The dissociation process necessarily includes a change of the kinetic energies of the fragments, away from the initial KE value of the parent molecular ion, that depends on the $E_{\text{KER}}$ value. It is thus a purely molecular effect and in addition to the KE changes that will experience the photoionised ion when travelling from the high-voltage biased interaction region to the field-free region preceding the entry into the magnetic analyser region. We discuss below these changes in the context of our experimental conditions. As we have just shown that the fragments can dissociate at any angle $\theta$ such that $0\leq \theta \leq\theta_{\text{max}}$ with $\sin\theta_{\text{max}}=v_0/V_C$, we calculate the possible values of the fragments $KE$ for the two possible extreme values of $\theta=0 \ \text{and}\ \theta =\theta_{\text{max}}$. This is done readily by substituting these values in eqn~\ref{speed} and converting to $KE$ values. At $\theta=0\degree$, we can have
 
 \begin{subequations}
 \label{KEThetaZero}
\begin{eqnarray}
&KE_{\ce{X}}=\frac{1}{2}m_{\ce{X}}(V_C-v_0^{\ce{X}})^2\, \text{, together with} \  \nonumber \\
&KE_{\ce{^{1}H}}=\frac{1}{2}m_{\ce{^{1}H}}(V_C+v_0^{\ce{^{1}H}})^2 \\
&\text{or} \nonumber \\
&KE_{\ce{X}}=\frac{1}{2}m_{\ce{X}}(V_C+v_0^{\ce{X}})^2\, \text{, together with} \  \nonumber \\ 
&KE_{\ce{^{1}H}}=\frac{1}{2}m_{\ce{^{1}H}}(V_C-v_0^{\ce{^{1}H}})^2.
\end{eqnarray}
\end{subequations}

\begin{figure}
\centering
  \includegraphics[width=8.2cm]{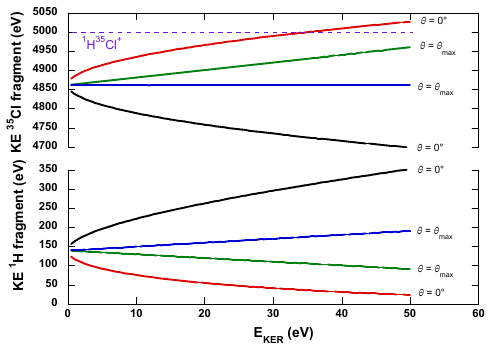}
  \caption{Kinetic energy (KE) of the \ce{^{35}Cl} (upper part) and \ce{^{1}H} (lower part) dissociation fragments as a function of $E_{\text{KER}}$ (eV) at $\theta=0\degree \ \text{and} \ \theta=\theta_{\text{max}}\degree$ dissociation angles. The purple coloured broken line at 5 keV represents the kinetic energy of the \ce{^1H^{35}Cl^{+}} parent molecular ion, ie. $V=1.637\times10^5$ ms$^{-1}$. Black solid lines: $KE_{\ce{^{35}Cl}}=\frac{1}{2}m_{\ce{^{35}Cl}}(V-v_0^{\ce{^{35}Cl}})^2$ and $KE_{\ce{^{1}H}}=\frac{1}{2}m_{\ce{^{1}H}}(V+v_0^{\ce{^{1}H}})^2$; red solid lines: $KE_{\ce{^{35}Cl}}=\frac{1}{2}m_{\ce{^{35}Cl}}(V+v_0^{\ce{^{35}Cl}})^2$ and $KE_{\ce{^{1}H}}=\frac{1}{2}m_{\ce{^{1}H}}(V-v_0^{\ce{^{1}H}})^2$, with $v_0^{\ce{^{35}Cl}}=391.1\sqrt{E_{\text{KER}}(\text{eV})}$ (ms$^{-1}$) and $v_0^{\ce{^{1}H}}=35v_0^{\ce{^{35}Cl}}$. Green solid lines: $KE_{\ce{^{1}H}}=\frac{1}{2}m_{\ce{^{1}H}}(V\cos \theta_{\text{max}}^{\ce{^{1}H}})^2$ with $\theta_{\text{max}}^{\ce{^{1}H}}$ shown as the \ce{^{1}H} black solid line on Fig.~\ref{Scattering angles}; see main text for calculation of $KE_{\ce{^{35}Cl}}$ in this case; blue solid lines: $KE_{\ce{^{35}Cl}}=\frac{1}{2}m_{\ce{^{35}Cl}}(V\cos \theta_{\text{max}}^{\ce{^{35}Cl}})^2$ with $\theta_{\text{max}}^{\ce{^{35}Cl}}$ shown as the \ce{^{35}Cl} black solid line on Fig.~\ref{Scattering angles}; see main text for calculation of $KE_{\ce{^{1}H}}$ in this case and additional background details.}  \label{KEFragments2}
\end{figure}

From Figs~\ref{KEFragments1} and~\ref{KEFragments2}, we see that the $KE$ of one of the fragments either increases or decreases when that of the other fragment conversely decreases or increases, as a function of increasing KER. Similar behaviour is observed when the fragments dissociate at their maximum angle $\theta_{\text{max}}$ (see solid lines in Fig.~\ref{Scattering angles}). In which case we have
\begin{subequations}
\label{KEThetaMax}
\begin{eqnarray}
&KE_{\ce{X}}=\frac{1}{2}m_{\ce{X}}(V_C\cos \theta_{\text{max}}^{\ce{X}})^2 \ \text{, together with} \ \nonumber \\
 &KE_{\ce{H}}=\frac{1}{2}m_{\ce{H}}\Big(V_C\cos \theta^{\ce{H}}+v_0^{\ce{H}}\sqrt{1-(V_C/v_0^{\ce{H}})^2\sin^2\theta^{\ce{H}}}\Big)^2 \label{KE1} \\
&\text{or} \nonumber\\ 
& KE_{\ce{H}}=\frac{1}{2}m_{\ce{H}}(V_C\cos \theta_{\text{max}}^{\ce{H}})^2 \ \text{, together with} \ \nonumber \\
&KE_{\ce{X}}=\frac{1}{2}m_{\ce{X}}\Big(V_C\cos \theta^{\ce{X}}+v_0^{\ce{H}}\sqrt{1-(V_C/v_0^{\ce{X}})^2\sin^2\theta^{\ce{X}}}\Big)^2. \label{KE2}
\end{eqnarray}
\end{subequations}
The positive root in eqn~\ref{speed} has been kept to obtain equations~\ref{KE1} and~\ref{KE2} to ensure adequate conservation of the total energy, i.e. $KE_{\ce{HX}}+E_{\text{KER}}=KE_{\ce{X}}+KE_{\ce{H}}$. The values of $\theta^{\ce{H}}$ and $\theta^{\ce{X}}$ to use in equations~\ref{KE1} and~\ref{KE2} are obtained from the \ce{^{1}H} and \ce{X} broken lines on Fig.~\ref{Scattering angles}, respectively. 

Focusing now on the heavy fragments in Figs.~\ref{KEFragments1} and ~\ref{KEFragments2}, we see that the effect of KER is to distribute homogeneously (quadratically from eqn~\ref{speed}) their kinetic energies between the minimum and maximum possible  values of $KE_{\ce{X}}=\frac{1}{2}m_{\ce{X}}(V_C-v_0^{\ce{X}})^2$ and $KE_{\ce{X}}=\frac{1}{2}m_{\ce{X}}(V_C+v_0^{\ce{X}})^2$, respectively, about the $KE_X$ value of $\frac{1}{2}m_{\ce{X}}V_C^2$. We take \ce{^{32}S} for a $E_{\text{KER}}$ value of 10 eV\footnotemark[5] \footnotetext[5]{It is not necessary to consider the width of the KER distribution~\citep{Chiang_2012} in this discussion.} as an example. Using equations~\ref{KEThetaZero} and ~\ref{KEThetaMax} and the data from Figure~\ref{Scattering angles} gives $KE_{\text{max}}$ and $KE_{\text{min}}$ values of $(4848 + 75)=4925$ eV and $(4848 - 75)=4773$ eV\footnotemark[3]\footnotetext[3]{Further analytical developments would show that the probability of dissociation with very low recoil angles is small and therefore the range of energies considered in the present example is certainly larger than the actual one. This further supports our conclusions.} in the interaction zone, respectively. Using equation~\ref{Magnet2} and assuming single ionisation,  KE values of $2925$ eV and $2773$ eV, after the tagging/interaction zone, would correspond to a magnetic field difference on the exit electro-magnet analyser of $\sim$23 G or, equivalently, a current difference of $\sim$ 0.7 A. We emphasise that this is an ideal, calculated value approximating the overall broadening effect of the KER on the ions final KE value prior to entering the analyser/demerger magnet. 

\begin{figure}
\centering
  \includegraphics[width=8.2cm]{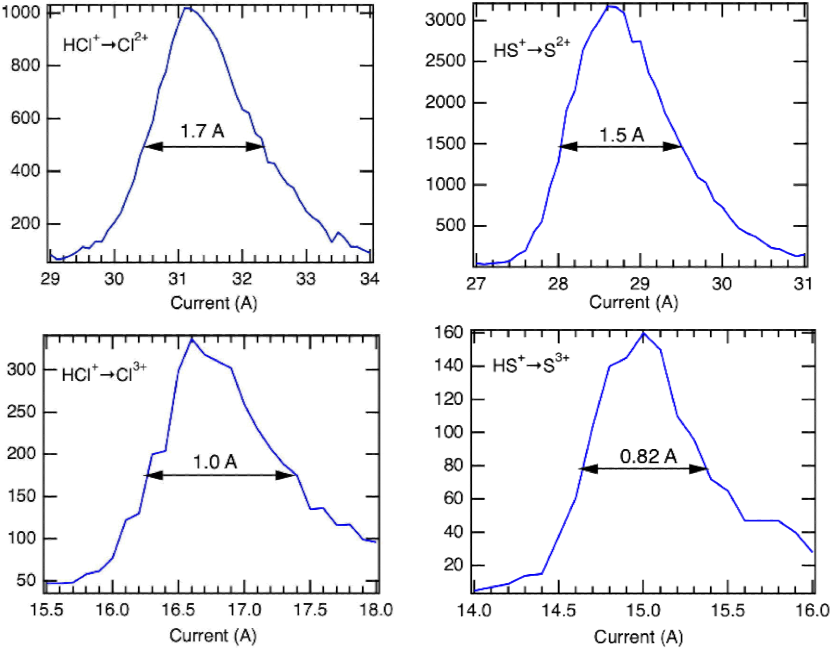}
  \caption{Current on demerger/analysing electro-magnet vs photo-ion yield signal (in arbitrary units) showing the optimisation of the four ionisation routes that were explored in this work: \ce{HS^+ ->S^{2+}} and \ce{HCl^+ ->Cl^{2+}}, i.e. the single ionisation channels, shown in the top half of the figure and \ce{HS^{+} -> S^{3+}} and \ce{HCl^+ ->Cl^{3+}}, i.e. the double ionisation channels, shown in the bottom half of the figure}
  \label{Magnet2Resolution}
\end{figure}

From figure~\ref{Magnet2Resolution}, we see that the value of 0.7 A is noticeably smaller than the measured resolution parameter of $\sim$ 1.5 A $\equiv$ 40 G $\equiv$ 260 eV FWHM of the analysing electro-magnet set for detection of the \ce{^{32}S^{2+}}ions. The measured widths of the curves of figure~\ref{Magnet2Resolution} comprise all the -always present- instrumental effects such the fragments refocalisation using ion optics, the fragments emittance at the interaction region exit and the aberrations of the ion optics, for the most part. These independent effects add in quadrature to the KER broadening contribution to produce the final width. Figure~\ref{Magnet2Resolution} shows this data for the four ionic fragmentation routes, i.e. four different demerger magnet settings, that were used in this work: $\ce{HS^+}\rightarrow \ce{S^{2+}} \text{ and } \ce{HCl^+}\rightarrow \ce{Cl^{2+}}$ -the single ionisation/fragmentation channels shown in the top half of Fig~\ref{Magnet2Resolution}- and $\ce{HS^+}\rightarrow \ce{S^{3+}} \text{ and } \ce{HCl^+}\rightarrow \ce{Cl^{3+}}$ -the double ionisation/fragmentation channels shown in the bottom half of Fig~\ref{Magnet2Resolution}-. Thus, we conclude that the bulk of the ion kinetic energies comprised between $KE_{\text{max}}$ and $KE_{\text{min}}$ is in all probability transmitted through to the detection stage for $E_{\text{KER}}$ values of up to a few ten's of eV. For larger $E_{\text{KER}}$ values, this experimental situation would likely become less well verified, resulting in some truncation of the spectrum of transmitted $KE_{\ce{X}}$ values. From Figs.~\ref{KEFragments2} and ~\ref{Magnet2Resolution}, similar conclusions hold for \ce{^{35}Cl}.

\section*{Acknowledgements}
The authors thank the SOLEIL PLEIADES beamline staff John Bozek, Aleksandar Milosavljevic, Christophe Nicolas and Emmanuel Robert for their help and interesting discussions during the experiments.





\bibliography{MainDocument} 

\providecommand*{\mcitethebibliography}{\thebibliography}
\csname @ifundefined\endcsname{endmcitethebibliography}
{\let\endmcitethebibliography\endthebibliography}{}
\begin{mcitethebibliography}{61}
\providecommand*{\natexlab}[1]{#1}
\providecommand*{\mciteSetBstSublistMode}[1]{}
\providecommand*{\mciteSetBstMaxWidthForm}[2]{}
\providecommand*{\mciteBstWouldAddEndPuncttrue}
  {\def\EndOfBibitem{\unskip.}}
\providecommand*{\mciteBstWouldAddEndPunctfalse}
  {\let\EndOfBibitem\relax}
\providecommand*{\mciteSetBstMidEndSepPunct}[3]{}
\providecommand*{\mciteSetBstSublistLabelBeginEnd}[3]{}
\providecommand*{\EndOfBibitem}{}
\mciteSetBstSublistMode{f}
\mciteSetBstMaxWidthForm{subitem}
{(\emph{\alph{mcitesubitemcount}})}
\mciteSetBstSublistLabelBeginEnd{\mcitemaxwidthsubitemform\space}
{\relax}{\relax}

\bibitem[Guha \emph{et~al.}(2008)Guha, Donnelly, and Pu]{Guha_2008}
J.~Guha, V.~M. Donnelly and Y.-K. Pu, \emph{Journal of Applied Physics}, 2008,
  \textbf{103}, 013306\relax
\mciteBstWouldAddEndPuncttrue
\mciteSetBstMidEndSepPunct{\mcitedefaultmidpunct}
{\mcitedefaultendpunct}{\mcitedefaultseppunct}\relax
\EndOfBibitem
\bibitem[{\'A}d{\'a}mkovics \emph{et~al.}(2011){\'A}d{\'a}mkovics, Glassgold,
  and Meijerink]{Adamkovics_2011}
M.~{\'A}d{\'a}mkovics, A.~E. Glassgold and R.~Meijerink, \emph{The
  Astrophysical Journal}, 2011, \textbf{736}, 143\relax
\mciteBstWouldAddEndPuncttrue
\mciteSetBstMidEndSepPunct{\mcitedefaultmidpunct}
{\mcitedefaultendpunct}{\mcitedefaultseppunct}\relax
\EndOfBibitem
\bibitem[Gatuzz \emph{et~al.}(2024)Gatuzz, Gorczyca, Hasoglu, Garc{\'\i}a, and
  Kallman]{Gatuzz_2024}
E.~Gatuzz, T.~W. Gorczyca, M.~F. Hasoglu, J.~A. Garc{\'\i}a and T.~R. Kallman,
  \emph{A\&A}, 2024, \textbf{689}, A325\relax
\mciteBstWouldAddEndPuncttrue
\mciteSetBstMidEndSepPunct{\mcitedefaultmidpunct}
{\mcitedefaultendpunct}{\mcitedefaultseppunct}\relax
\EndOfBibitem
\bibitem[Howell(2008)]{Howell_2008}
R.~W. Howell, \emph{International Journal of Radiation Biology}, 2008,
  \textbf{84}, 959--975\relax
\mciteBstWouldAddEndPuncttrue
\mciteSetBstMidEndSepPunct{\mcitedefaultmidpunct}
{\mcitedefaultendpunct}{\mcitedefaultseppunct}\relax
\EndOfBibitem
\bibitem[Hofmann(2014)]{Hofmann_2014}
S.~Hofmann, \emph{Auger- and X-Ray Photoelectron Spectroscopy in Materials
  Science: A User-Oriented Guide}, Springer Berlin Heidelberg, 2014\relax
\mciteBstWouldAddEndPuncttrue
\mciteSetBstMidEndSepPunct{\mcitedefaultmidpunct}
{\mcitedefaultendpunct}{\mcitedefaultseppunct}\relax
\EndOfBibitem
\bibitem[Schippers and M{\"u}ller(2020)]{Schippers_2020}
S.~Schippers and A.~M{\"u}ller, \emph{Atoms}, 2020, \textbf{8}, 45\relax
\mciteBstWouldAddEndPuncttrue
\mciteSetBstMidEndSepPunct{\mcitedefaultmidpunct}
{\mcitedefaultendpunct}{\mcitedefaultseppunct}\relax
\EndOfBibitem
\bibitem[Kennedy(2022)]{Kennedy_2022}
E.~T. Kennedy, \emph{Physica Scripta}, 2022, \textbf{97}, 054003\relax
\mciteBstWouldAddEndPuncttrue
\mciteSetBstMidEndSepPunct{\mcitedefaultmidpunct}
{\mcitedefaultendpunct}{\mcitedefaultseppunct}\relax
\EndOfBibitem
\bibitem[Dehmer(1972)]{Dehmer_1972}
J.~L. Dehmer, \emph{The Journal of Chemical Physics}, 1972, \textbf{56},
  4496--4504\relax
\mciteBstWouldAddEndPuncttrue
\mciteSetBstMidEndSepPunct{\mcitedefaultmidpunct}
{\mcitedefaultendpunct}{\mcitedefaultseppunct}\relax
\EndOfBibitem
\bibitem[Lindblad \emph{et~al.}(2020)Lindblad, Kjellsson, Couto, Timm,
  B{\"u}low, Zamudio-Bayer, Lundberg, von Issendorff, Lau, Sorensen,
  Carravetta, {\AA}gren, and Rubensson]{Lindblad_2020}
R.~Lindblad, L.~Kjellsson, R.~C. Couto, M.~Timm, C.~B{\"u}low,
  V.~Zamudio-Bayer, M.~Lundberg, B.~von Issendorff, J.~T. Lau, S.~L. Sorensen,
  V.~Carravetta, H.~{\AA}gren and J.~E. Rubensson, \emph{Physical Review
  Letters}, 2020, \textbf{124}, 203001--\relax
\mciteBstWouldAddEndPuncttrue
\mciteSetBstMidEndSepPunct{\mcitedefaultmidpunct}
{\mcitedefaultendpunct}{\mcitedefaultseppunct}\relax
\EndOfBibitem
\bibitem[Lindblad \emph{et~al.}(2022)Lindblad, Kjellsson, De~Santis,
  Zamudio-Bayer, von Issendorff, Sorensen, Lau, Hua, Carravetta, Rubensson,
  {\AA}gren, and Couto]{Lindblad_2022}
R.~Lindblad, L.~Kjellsson, E.~De~Santis, V.~Zamudio-Bayer, B.~von Issendorff,
  S.~L. Sorensen, J.~T. Lau, W.~Hua, V.~Carravetta, J.-E. Rubensson,
  H.~{\AA}gren and R.~C. Couto, \emph{Physical Review A}, 2022, \textbf{106},
  042814--\relax
\mciteBstWouldAddEndPuncttrue
\mciteSetBstMidEndSepPunct{\mcitedefaultmidpunct}
{\mcitedefaultendpunct}{\mcitedefaultseppunct}\relax
\EndOfBibitem
\bibitem[Schwarz \emph{et~al.}(2022)Schwarz, Kielgast, Baev, Reinwardt,
  Trinter, Klumpp, Perry-Sassmannshausen, Buhr, Schippers, M{\"u}ller, Bari,
  Mondes, Flesch, R{\"u}hl, and Martins]{Schwarz_2022}
J.~Schwarz, F.~Kielgast, I.~Baev, S.~Reinwardt, F.~Trinter, S.~Klumpp,
  A.~Perry-Sassmannshausen, T.~Buhr, S.~Schippers, A.~M{\"u}ller, S.~Bari,
  V.~Mondes, R.~Flesch, E.~R{\"u}hl and M.~Martins, \emph{Physical Chemistry
  Chemical Physics}, 2022, \textbf{24}, 23119--23127\relax
\mciteBstWouldAddEndPuncttrue
\mciteSetBstMidEndSepPunct{\mcitedefaultmidpunct}
{\mcitedefaultendpunct}{\mcitedefaultseppunct}\relax
\EndOfBibitem
\bibitem[Reinwardt \emph{et~al.}(2024)Reinwardt, Cieslik, Buhr,
  Perry-Sassmannshausen, Schippers, M{\"u}ller, Trinter, and
  Martins]{Reinwardt_2024}
S.~Reinwardt, P.~Cieslik, T.~Buhr, A.~Perry-Sassmannshausen, S.~Schippers,
  A.~M{\"u}ller, F.~Trinter and M.~Martins, \emph{Phys. Chem. Chem. Phys.},
  2024, \textbf{26}, 15519--15529\relax
\mciteBstWouldAddEndPuncttrue
\mciteSetBstMidEndSepPunct{\mcitedefaultmidpunct}
{\mcitedefaultendpunct}{\mcitedefaultseppunct}\relax
\EndOfBibitem
\bibitem[Cornetta \emph{et~al.}(2025)Cornetta, Kjellsson, Couto, {\AA}gren,
  Carravetta, S{\"o}rensen, Kubin, B{\"u}low, Zamudio-Bayer, von Issendorff,
  Lau, S{\"o}derstr{\"o}m, Ag{\aa}ker, Rubensson, and Lindblad]{Cornetta_2025}
L.~M. Cornetta, L.~Kjellsson, R.~C. Couto, H.~{\AA}gren, V.~Carravetta, S.~L.
  S{\"o}rensen, M.~Kubin, C.~B{\"u}low, V.~Zamudio-Bayer, B.~von Issendorff,
  J.~T. Lau, J.~S{\"o}derstr{\"o}m, M.~Ag{\aa}ker, J.-E. Rubensson and
  R.~Lindblad, \emph{Physical Review A}, 2025, \textbf{111}, 022808--\relax
\mciteBstWouldAddEndPuncttrue
\mciteSetBstMidEndSepPunct{\mcitedefaultmidpunct}
{\mcitedefaultendpunct}{\mcitedefaultseppunct}\relax
\EndOfBibitem
\bibitem[Mosnier \emph{et~al.}(2016)Mosnier, Kennedy, van Kampen, Cubaynes,
  Guilbaud, Sisourat, Puglisi, Carniato, and Bizau]{Mosnier_2016}
J.~P. Mosnier, E.~T. Kennedy, P.~van Kampen, D.~Cubaynes, S.~Guilbaud,
  N.~Sisourat, A.~Puglisi, S.~Carniato and J.~M. Bizau, \emph{Physical Review
  A}, 2016, \textbf{93}, 061401--\relax
\mciteBstWouldAddEndPuncttrue
\mciteSetBstMidEndSepPunct{\mcitedefaultmidpunct}
{\mcitedefaultendpunct}{\mcitedefaultseppunct}\relax
\EndOfBibitem
\bibitem[Klumpp \emph{et~al.}(2018)Klumpp, Guda, Schubert, Mertens, Hellhund,
  M{\"u}ller, Schippers, Bari, and Martins]{Klumpp_2018}
S.~Klumpp, A.~A. Guda, K.~Schubert, K.~Mertens, J.~Hellhund, A.~M{\"u}ller,
  S.~Schippers, S.~Bari and M.~Martins, \emph{Physical Review A}, 2018,
  \textbf{97}, 033401--\relax
\mciteBstWouldAddEndPuncttrue
\mciteSetBstMidEndSepPunct{\mcitedefaultmidpunct}
{\mcitedefaultendpunct}{\mcitedefaultseppunct}\relax
\EndOfBibitem
\bibitem[Kennedy \emph{et~al.}(2018)Kennedy, Mosnier, van Kampen, Bizau,
  Cubaynes, Guilbaud, Carniato, Puglisi, and Sisourat]{Kennedy_2018}
E.~T. Kennedy, J.~P. Mosnier, P.~van Kampen, J.~M. Bizau, D.~Cubaynes,
  S.~Guilbaud, S.~Carniato, A.~Puglisi and N.~Sisourat, \emph{Physical Review
  A}, 2018, \textbf{97}, 043410--\relax
\mciteBstWouldAddEndPuncttrue
\mciteSetBstMidEndSepPunct{\mcitedefaultmidpunct}
{\mcitedefaultendpunct}{\mcitedefaultseppunct}\relax
\EndOfBibitem
\bibitem[Bari \emph{et~al.}(2019)Bari, Inhester, Schubert, Mertens, Schunck,
  D{\"o}rner, Deinert, Schwob, Schippers, M{\"u}ller, Klumpp, and
  Martins]{Bari_2019}
S.~Bari, L.~Inhester, K.~Schubert, K.~Mertens, J.~O. Schunck, S.~D{\"o}rner,
  S.~Deinert, L.~Schwob, S.~Schippers, A.~M{\"u}ller, S.~Klumpp and M.~Martins,
  \emph{Physical Chemistry Chemical Physics}, 2019, \textbf{21},
  16505--16514\relax
\mciteBstWouldAddEndPuncttrue
\mciteSetBstMidEndSepPunct{\mcitedefaultmidpunct}
{\mcitedefaultendpunct}{\mcitedefaultseppunct}\relax
\EndOfBibitem
\bibitem[Kobayashi \emph{et~al.}(2019)Kobayashi, Zeng, Neumark, and
  Leone]{Kobayashi_2019}
Y.~Kobayashi, T.~Zeng, D.~M. Neumark and S.~R. Leone, \emph{Structural
  Dynamics}, 2019, \textbf{6}, 014101\relax
\mciteBstWouldAddEndPuncttrue
\mciteSetBstMidEndSepPunct{\mcitedefaultmidpunct}
{\mcitedefaultendpunct}{\mcitedefaultseppunct}\relax
\EndOfBibitem
\bibitem[Carniato \emph{et~al.}(2020)Carniato, Bizau, Cubaynes, Kennedy,
  Guilbaud, Sokell, McLaughlin, and Mosnier]{Carniato_2020}
S.~Carniato, J.-M. Bizau, D.~Cubaynes, E.~T. Kennedy, S.~Guilbaud, E.~Sokell,
  B.~McLaughlin and J.-P. Mosnier, \emph{Atoms}, 2020, \textbf{8}, 67\relax
\mciteBstWouldAddEndPuncttrue
\mciteSetBstMidEndSepPunct{\mcitedefaultmidpunct}
{\mcitedefaultendpunct}{\mcitedefaultseppunct}\relax
\EndOfBibitem
\bibitem[Martins \emph{et~al.}(2021)Martins, Reinwardt, Schunck, Schwarz, Baev,
  M{\"u}ller, Buhr, Perry-Sassmannshausen, Klumpp, and Schippers]{Martins_2021}
M.~Martins, S.~Reinwardt, J.~O. Schunck, J.~Schwarz, K.~Baev, A.~M{\"u}ller,
  T.~Buhr, A.~Perry-Sassmannshausen, S.~Klumpp and S.~Schippers, \emph{The
  Journal of Physical Chemistry Letters}, 2021, \textbf{12}, 1390--1395\relax
\mciteBstWouldAddEndPuncttrue
\mciteSetBstMidEndSepPunct{\mcitedefaultmidpunct}
{\mcitedefaultendpunct}{\mcitedefaultseppunct}\relax
\EndOfBibitem
\bibitem[Puglisi \emph{et~al.}(2018)Puglisi, Miteva, Kennedy, Mosnier, Bizau,
  Cubaynes, Sisourat, and Carniato]{Puglisi_2018}
A.~Puglisi, T.~Miteva, E.~T. Kennedy, J.-P. Mosnier, J.-M. Bizau, D.~Cubaynes,
  N.~Sisourat and S.~Carniato, \emph{Physical Chemistry Chemical Physics},
  2018, \textbf{20}, 4415--4421\relax
\mciteBstWouldAddEndPuncttrue
\mciteSetBstMidEndSepPunct{\mcitedefaultmidpunct}
{\mcitedefaultendpunct}{\mcitedefaultseppunct}\relax
\EndOfBibitem
\bibitem[Walsh \emph{et~al.}(2012)Walsh, Nomura, Millar, and
  Aikawa]{Walsh_2012}
C.~Walsh, H.~Nomura, T.~J. Millar and Y.~Aikawa, \emph{The Astrophysical
  Journal}, 2012, \textbf{747}, 114\relax
\mciteBstWouldAddEndPuncttrue
\mciteSetBstMidEndSepPunct{\mcitedefaultmidpunct}
{\mcitedefaultendpunct}{\mcitedefaultseppunct}\relax
\EndOfBibitem
\bibitem[Abel \emph{et~al.}(2008)Abel, Federman, and Stancil]{Abel_2008}
N.~P. Abel, S.~R. Federman and P.~C. Stancil, \emph{The Astrophysical Journal},
  2008, \textbf{675}, L81--L84\relax
\mciteBstWouldAddEndPuncttrue
\mciteSetBstMidEndSepPunct{\mcitedefaultmidpunct}
{\mcitedefaultendpunct}{\mcitedefaultseppunct}\relax
\EndOfBibitem
\bibitem[St{\"a}uber \emph{et~al.}(2005)St{\"a}uber, Doty, van Dishoeck, and
  Benz]{Stauber_2005}
P.~St{\"a}uber, S.~D. Doty, E.~F. van Dishoeck and A.~O. Benz, \emph{A\&A},
  2005, \textbf{440}, 949--966\relax
\mciteBstWouldAddEndPuncttrue
\mciteSetBstMidEndSepPunct{\mcitedefaultmidpunct}
{\mcitedefaultendpunct}{\mcitedefaultseppunct}\relax
\EndOfBibitem
\bibitem[Benz \emph{et~al.}(2010)Benz, Bruderer, van Dishoeck, St{\"a}uber,
  Wampfler, Melchior, Dedes, Wyrowski, Doty, van~der Tak, B{\"a}chtold,
  Csillaghy, Megej, Monstein, Soldati, Bachiller, Baudry, Benedettini, Bergin,
  Bjerkeli, Blake, Bontemps, Braine, Caselli, Cernicharo, Codella, Daniel,
  di~Giorgio, Dieleman, Dominik, Encrenaz, Fich, Fuente, Giannini, Goicoechea,
  de~Graauw, Helmich, Herczeg, Herpin, Hogerheijde, Jacq, Jellema, Johnstone,
  J{\o}rgensen, Kristensen, Larsson, Lis, Liseau, Marseille, McCoey, Melnick,
  Neufeld, Nisini, Olberg, Ossenkopf, Parise, Pearson, Plume, Risacher,
  Santiago-Garc{\'\i}a, Saraceno, Schieder, Shipman, Stutzki, Tafalla, Tielens,
  van Kempen, Visser, and Yıldız]{Benz_2010}
A.~O. Benz, S.~Bruderer, E.~F. van Dishoeck, P.~St{\"a}uber, S.~F. Wampfler,
  M.~Melchior, C.~Dedes, F.~Wyrowski, S.~D. Doty, F.~van~der Tak,
  W.~B{\"a}chtold, A.~Csillaghy, A.~Megej, C.~Monstein, M.~Soldati,
  R.~Bachiller, A.~Baudry, M.~Benedettini, E.~Bergin, P.~Bjerkeli, G.~A. Blake,
  S.~Bontemps, J.~Braine, P.~Caselli, J.~Cernicharo, C.~Codella, F.~Daniel,
  A.~M. di~Giorgio, P.~Dieleman, C.~Dominik, P.~Encrenaz, M.~Fich, A.~Fuente,
  T.~Giannini, J.~R. Goicoechea, T.~de~Graauw, F.~Helmich, G.~J. Herczeg,
  F.~Herpin, M.~R. Hogerheijde, T.~Jacq, W.~Jellema, D.~Johnstone, J.~K.
  J{\o}rgensen, L.~E. Kristensen, B.~Larsson, D.~Lis, R.~Liseau, M.~Marseille,
  C.~McCoey, G.~Melnick, D.~Neufeld, B.~Nisini, M.~Olberg, V.~Ossenkopf,
  B.~Parise, J.~C. Pearson, R.~Plume, C.~Risacher, J.~Santiago-Garc{\'\i}a,
  P.~Saraceno, R.~Schieder, R.~Shipman, J.~Stutzki, M.~Tafalla, A.~G. G.~M.
  Tielens, T.~A. van Kempen, R.~Visser and U.~A. Yıldız, \emph{A\&A}, 2010,
  \textbf{521}, L35\relax
\mciteBstWouldAddEndPuncttrue
\mciteSetBstMidEndSepPunct{\mcitedefaultmidpunct}
{\mcitedefaultendpunct}{\mcitedefaultseppunct}\relax
\EndOfBibitem
\bibitem[Goicoechea \emph{et~al.}(2021)Goicoechea, Aguado, Cuadrado, Roncero,
  Pety, Bron, Fuente, Riquelme, Chapillon, Herrera, and Duran]{Goicoechea_2021}
J.~R. Goicoechea, A.~Aguado, S.~Cuadrado, O.~Roncero, J.~Pety, E.~Bron,
  A.~Fuente, D.~Riquelme, E.~Chapillon, C.~Herrera and C.~A. Duran,
  \emph{A\&A}, 2021, \textbf{647}, A10\relax
\mciteBstWouldAddEndPuncttrue
\mciteSetBstMidEndSepPunct{\mcitedefaultmidpunct}
{\mcitedefaultendpunct}{\mcitedefaultseppunct}\relax
\EndOfBibitem
\bibitem[Gerin \emph{et~al.}(2016)Gerin, Neufeld, and Goicoechea]{Gerin_2016}
M.~Gerin, D.~A. Neufeld and J.~R. Goicoechea, \emph{Annual Review of Astronomy
  and Astrophysics}, 2016, \textbf{54}, 181--225\relax
\mciteBstWouldAddEndPuncttrue
\mciteSetBstMidEndSepPunct{\mcitedefaultmidpunct}
{\mcitedefaultendpunct}{\mcitedefaultseppunct}\relax
\EndOfBibitem
\bibitem[{Menten, K. M.} \emph{et~al.}(2011){Menten, K. M.}, {Wyrowski, F.},
  {Belloche, A.}, {G{\"u}sten, R.}, {Dedes, L.}, and {M{\"u}ller, H. S.
  P.}]{Menten_2011}
{Menten, K. M.}, {Wyrowski, F.}, {Belloche, A.}, {G{\"u}sten, R.}, {Dedes, L.}
  and {M{\"u}ller, H. S. P.}, \emph{A\&A}, 2011, \textbf{525}, A77\relax
\mciteBstWouldAddEndPuncttrue
\mciteSetBstMidEndSepPunct{\mcitedefaultmidpunct}
{\mcitedefaultendpunct}{\mcitedefaultseppunct}\relax
\EndOfBibitem
\bibitem[De~Luca \emph{et~al.}(2012)De~Luca, Gupta, Neufeld, Gerin, Teyssier,
  Drouin, Pearson, Lis, Monje, Phillips, Goicoechea, Godard, Falgarone,
  Coutens, and Bell]{De-Luca_2012}
M.~De~Luca, H.~Gupta, D.~Neufeld, M.~Gerin, D.~Teyssier, B.~J. Drouin, J.~C.
  Pearson, D.~C. Lis, R.~Monje, T.~G. Phillips, J.~R. Goicoechea, B.~Godard,
  E.~Falgarone, A.~Coutens and T.~A. Bell, \emph{The Astrophysical Journal
  Letters}, 2012, \textbf{751}, L37\relax
\mciteBstWouldAddEndPuncttrue
\mciteSetBstMidEndSepPunct{\mcitedefaultmidpunct}
{\mcitedefaultendpunct}{\mcitedefaultseppunct}\relax
\EndOfBibitem
\bibitem[Chiang \emph{et~al.}(2012)Chiang, Otto, Meyer, and
  Cederbaum]{Chiang_2012}
Y.-C. Chiang, F.~Otto, H.-D. Meyer and L.~S. Cederbaum, \emph{The Journal of
  Chemical Physics}, 2012, \textbf{136}, 114111\relax
\mciteBstWouldAddEndPuncttrue
\mciteSetBstMidEndSepPunct{\mcitedefaultmidpunct}
{\mcitedefaultendpunct}{\mcitedefaultseppunct}\relax
\EndOfBibitem
\bibitem[Carlson and White(1966)]{Carlson_1966}
T.~A. Carlson and R.~M. White, \emph{The Journal of Chemical Physics}, 1966,
  \textbf{44}, 4510--4520\relax
\mciteBstWouldAddEndPuncttrue
\mciteSetBstMidEndSepPunct{\mcitedefaultmidpunct}
{\mcitedefaultendpunct}{\mcitedefaultseppunct}\relax
\EndOfBibitem
\bibitem[Hitchcock \emph{et~al.}(1988)Hitchcock, Lablanquie, Morin, Lizon
  A~Lugrin, Simon, Thiry, and Nenner]{Hitchcock_1988}
A.~P. Hitchcock, P.~Lablanquie, P.~Morin, E.~Lizon A~Lugrin, M.~Simon, P.~Thiry
  and I.~Nenner, \emph{Physical Review A}, 1988, \textbf{37}, 2448--2466\relax
\mciteBstWouldAddEndPuncttrue
\mciteSetBstMidEndSepPunct{\mcitedefaultmidpunct}
{\mcitedefaultendpunct}{\mcitedefaultseppunct}\relax
\EndOfBibitem
\bibitem[Nenner and Morin(1996)]{Nenner_1996}
I.~Nenner and P.~Morin, in \emph{Electronic and Nuclear Relaxation Of
  Core-Excited Molecules}, ed. U.~Becker and D.~A. Shirley, Springer US,
  Boston, MA, 1996, pp. 291--354\relax
\mciteBstWouldAddEndPuncttrue
\mciteSetBstMidEndSepPunct{\mcitedefaultmidpunct}
{\mcitedefaultendpunct}{\mcitedefaultseppunct}\relax
\EndOfBibitem
\bibitem[Ueda and Eland(2005)]{Ueda_2005}
K.~Ueda and J.~H.~D. Eland, \emph{Journal of Physics B: Atomic, Molecular and
  Optical Physics}, 2005, \textbf{38}, S839\relax
\mciteBstWouldAddEndPuncttrue
\mciteSetBstMidEndSepPunct{\mcitedefaultmidpunct}
{\mcitedefaultendpunct}{\mcitedefaultseppunct}\relax
\EndOfBibitem
\bibitem[Falcinelli \emph{et~al.}(2014)Falcinelli, Rosi, Candori,
  Vecchiocattivi, Farrar, Pirani, Balucani, Alagia, Richter, and
  Stranges]{Falcinelli_2014}
S.~Falcinelli, M.~Rosi, P.~Candori, F.~Vecchiocattivi, J.~M. Farrar, F.~Pirani,
  N.~Balucani, M.~Alagia, R.~Richter and S.~Stranges, \emph{Planetary and Space
  Science}, 2014, \textbf{99}, 149--157\relax
\mciteBstWouldAddEndPuncttrue
\mciteSetBstMidEndSepPunct{\mcitedefaultmidpunct}
{\mcitedefaultendpunct}{\mcitedefaultseppunct}\relax
\EndOfBibitem
\bibitem[Scarlett \emph{et~al.}(2017)Scarlett, Zammit, Fursa, and
  Bray]{Scarlett_2017}
L.~H. Scarlett, M.~C. Zammit, D.~V. Fursa and I.~Bray, \emph{Physical Review
  A}, 2017, \textbf{96}, 022706--\relax
\mciteBstWouldAddEndPuncttrue
\mciteSetBstMidEndSepPunct{\mcitedefaultmidpunct}
{\mcitedefaultendpunct}{\mcitedefaultseppunct}\relax
\EndOfBibitem
\bibitem[Piancastelli \emph{et~al.}(2017)Piancastelli, Guillemin, Marchenko,
  Journel, Travnikova, Marin, Goldsztejn, de~Miranda, Ismail, and
  Simon]{Piancastelli_2017}
M.~N. Piancastelli, R.~Guillemin, T.~Marchenko, L.~Journel, O.~Travnikova,
  T.~Marin, G.~Goldsztejn, B.~C. de~Miranda, I.~Ismail and M.~Simon,
  \emph{Journal of Physics B: Atomic, Molecular and Optical Physics}, 2017,
  \textbf{50}, 042001\relax
\mciteBstWouldAddEndPuncttrue
\mciteSetBstMidEndSepPunct{\mcitedefaultmidpunct}
{\mcitedefaultendpunct}{\mcitedefaultseppunct}\relax
\EndOfBibitem
\bibitem[Rebholz \emph{et~al.}(2021)Rebholz, Ding, Aufleger, Hartmann, Meyer,
  Stoo{\ss}, Magunia, Wachs, Birk, Mi, Borisova, da~Costa~Castanheira,
  Rupprecht, Magrakvelidze, Thumm, Roling, Butz, Zacharias, D{\"u}sterer,
  Treusch, Brenner, Ott, and Pfeifer]{Rebholz_2021}
M.~Rebholz, T.~Ding, L.~Aufleger, M.~Hartmann, K.~Meyer, V.~Stoo{\ss},
  A.~Magunia, D.~Wachs, P.~Birk, Y.~Mi, G.~D. Borisova,
  C.~da~Costa~Castanheira, P.~Rupprecht, M.~Magrakvelidze, U.~Thumm, S.~Roling,
  M.~Butz, H.~Zacharias, S.~D{\"u}sterer, R.~Treusch, G.~Brenner, C.~Ott and
  T.~Pfeifer, \emph{The Journal of Physical Chemistry A}, 2021, \textbf{125},
  10138--10143\relax
\mciteBstWouldAddEndPuncttrue
\mciteSetBstMidEndSepPunct{\mcitedefaultmidpunct}
{\mcitedefaultendpunct}{\mcitedefaultseppunct}\relax
\EndOfBibitem
\bibitem[Bizau \emph{et~al.}(2016)Bizau, Cubaynes, Guilbaud, Eassan, Shorman,
  Bouisset, Guigand, Moustier, Mari{\'{e}}, Nadal, Robert, Nicolas, and
  Miron]{Bizau_2016}
J.~Bizau, D.~Cubaynes, S.~Guilbaud, N.~E. Eassan, M.~A. Shorman, E.~Bouisset,
  J.~Guigand, O.~Moustier, A.~Mari{\'{e}}, E.~Nadal, E.~Robert, C.~Nicolas and
  C.~Miron, \emph{Journal of Electron Spectroscopy and Related Phenomena},
  2016, \textbf{210}, 5--12\relax
\mciteBstWouldAddEndPuncttrue
\mciteSetBstMidEndSepPunct{\mcitedefaultmidpunct}
{\mcitedefaultendpunct}{\mcitedefaultseppunct}\relax
\EndOfBibitem
\bibitem[Mosnier \emph{et~al.}(2025)Mosnier, Kennedy, Cubaynes, Bizau,
  Guilbaud, Blancard, McLaughlin, Haso{\u g}lu, and Gorczyca]{Mosnier_2025}
J.-P. Mosnier, E.~T. Kennedy, D.~Cubaynes, J.-M. Bizau, S.~Guilbaud,
  C.~Blancard, B.~M. McLaughlin, M.~F. Haso{\u g}lu and T.~W. Gorczyca,
  \emph{Journal of Physics B: Atomic, Molecular and Optical Physics}, 2025,
  \textbf{58}, 075002\relax
\mciteBstWouldAddEndPuncttrue
\mciteSetBstMidEndSepPunct{\mcitedefaultmidpunct}
{\mcitedefaultendpunct}{\mcitedefaultseppunct}\relax
\EndOfBibitem
\bibitem[Ren \emph{et~al.}(2011)Ren, Wang, Li, Yuan, and Zhu]{Ren_2011}
L.-M. Ren, Y.-Y. Wang, D.-D. Li, Z.-S. Yuan and L.-F. Zhu, \emph{Chinese
  Physics Letters}, 2011, \textbf{28}, 053401\relax
\mciteBstWouldAddEndPuncttrue
\mciteSetBstMidEndSepPunct{\mcitedefaultmidpunct}
{\mcitedefaultendpunct}{\mcitedefaultseppunct}\relax
\EndOfBibitem
\bibitem[et~al(2020)]{Mosnier_2020}
J.~M. et~al, \emph{Photoionisation of chlorine ions in the 2p region},
  Unpublished, 2020, SOLEIL work\relax
\mciteBstWouldAddEndPuncttrue
\mciteSetBstMidEndSepPunct{\mcitedefaultmidpunct}
{\mcitedefaultendpunct}{\mcitedefaultseppunct}\relax
\EndOfBibitem
\bibitem[Schmidt \emph{et~al.}(1993)Schmidt, Baldridge, Boatz, Elbert, Gordon,
  Jensen, Koseki, Matsunaga, Nguyen, Su, Windus, Dupuis, and
  Montgomery]{GAMESS93}
M.~W. Schmidt, K.~K. Baldridge, J.~A. Boatz, S.~T. Elbert, M.~S. Gordon, J.~H.
  Jensen, S.~Koseki, N.~Matsunaga, K.~Nguyen, S.~Su, T.~Windus, M.~Dupuis and
  J.~Montgomery, \emph{J. Comput. Chem.}, 1993, \textbf{14}, 1347\relax
\mciteBstWouldAddEndPuncttrue
\mciteSetBstMidEndSepPunct{\mcitedefaultmidpunct}
{\mcitedefaultendpunct}{\mcitedefaultseppunct}\relax
\EndOfBibitem
\bibitem[Becke(1993)]{becke3}
A.~Becke, \emph{J. Chem. Phys.}, 1993, \textbf{98}, 5648--5652\relax
\mciteBstWouldAddEndPuncttrue
\mciteSetBstMidEndSepPunct{\mcitedefaultmidpunct}
{\mcitedefaultendpunct}{\mcitedefaultseppunct}\relax
\EndOfBibitem
\bibitem[Lee \emph{et~al.}(1998)Lee, Yang, and Parr]{lyp}
C.~Lee, W.~Yang and R.~Parr, \emph{Phys. Rev. B}, 1998, \textbf{38}, 3098\relax
\mciteBstWouldAddEndPuncttrue
\mciteSetBstMidEndSepPunct{\mcitedefaultmidpunct}
{\mcitedefaultendpunct}{\mcitedefaultseppunct}\relax
\EndOfBibitem
\bibitem[Nakajima and Hirao(2003)]{DK3}
T.~Nakajima and K.~Hirao, \emph{J. Chem. Phys.}, 2003, \textbf{119},
  4105--4111\relax
\mciteBstWouldAddEndPuncttrue
\mciteSetBstMidEndSepPunct{\mcitedefaultmidpunct}
{\mcitedefaultendpunct}{\mcitedefaultseppunct}\relax
\EndOfBibitem
\bibitem[Woon and Dunning(1994)]{Woon1994}
D.~E. Woon and T.~H. Dunning, \emph{Journal of Chemical Physics}, 1994,
  \textbf{100}, 2975--2988\relax
\mciteBstWouldAddEndPuncttrue
\mciteSetBstMidEndSepPunct{\mcitedefaultmidpunct}
{\mcitedefaultendpunct}{\mcitedefaultseppunct}\relax
\EndOfBibitem
\bibitem[Pritchard \emph{et~al.}(2019)Pritchard, Altarawy, Didier, Gibson, and
  Windus]{Pritchard2019a}
B.~P. Pritchard, D.~Altarawy, B.~Didier, T.~D. Gibson and T.~L. Windus,
  \emph{Journal of Chemical Information and Modeling}, 2019, \textbf{59},
  4814--4820\relax
\mciteBstWouldAddEndPuncttrue
\mciteSetBstMidEndSepPunct{\mcitedefaultmidpunct}
{\mcitedefaultendpunct}{\mcitedefaultseppunct}\relax
\EndOfBibitem
\bibitem[Journel \emph{et~al.}(2008)Journel, Guillemin, Haouas, Lablanquie,
  Penent, Palaudoux, Andric, Simon, C{\'e}olin, Kaneyasu, Viefhaus, Braune, Li,
  Elkharrat, Catoire, Houver, and Dowek]{Journel_2008}
L.~Journel, R.~Guillemin, A.~Haouas, P.~Lablanquie, F.~Penent, J.~Palaudoux,
  L.~Andric, M.~Simon, D.~C{\'e}olin, T.~Kaneyasu, J.~Viefhaus, M.~Braune,
  W.~B. Li, C.~Elkharrat, F.~Catoire, J.~C. Houver and D.~Dowek, \emph{Physical
  Review A}, 2008, \textbf{77}, 042710--\relax
\mciteBstWouldAddEndPuncttrue
\mciteSetBstMidEndSepPunct{\mcitedefaultmidpunct}
{\mcitedefaultendpunct}{\mcitedefaultseppunct}\relax
\EndOfBibitem
\bibitem[Hudson \emph{et~al.}(1994)Hudson, Shirley, Domke, Remmers, and
  Kaindl]{Hudson_1994}
E.~Hudson, D.~A. Shirley, M.~Domke, G.~Remmers and G.~Kaindl, \emph{Physical
  Review A}, 1994, \textbf{49}, 161--175\relax
\mciteBstWouldAddEndPuncttrue
\mciteSetBstMidEndSepPunct{\mcitedefaultmidpunct}
{\mcitedefaultendpunct}{\mcitedefaultseppunct}\relax
\EndOfBibitem
\bibitem[Aksela \emph{et~al.}(1992)Aksela, Aksela, Sairanen, Kivim{\"a}ki,
  Bancroft, and Tan]{H_Aksela_1992a}
H.~Aksela, S.~Aksela, O.-P. Sairanen, A.~Kivim{\"a}ki, G.~M. Bancroft and K.~H.
  Tan, \emph{Physica Scripta}, 1992, \textbf{1992}, 122\relax
\mciteBstWouldAddEndPuncttrue
\mciteSetBstMidEndSepPunct{\mcitedefaultmidpunct}
{\mcitedefaultendpunct}{\mcitedefaultseppunct}\relax
\EndOfBibitem
\bibitem[Aksela \emph{et~al.}(1992)Aksela, Aksela, Naves~de Brito, Bancroft,
  and Tan]{H_Aksela_1992c}
H.~Aksela, S.~Aksela, A.~Naves~de Brito, G.~M. Bancroft and K.~H. Tan,
  \emph{Physical Review A}, 1992, \textbf{45}, 7948--7952\relax
\mciteBstWouldAddEndPuncttrue
\mciteSetBstMidEndSepPunct{\mcitedefaultmidpunct}
{\mcitedefaultendpunct}{\mcitedefaultseppunct}\relax
\EndOfBibitem
\bibitem[Le~Guen \emph{et~al.}(2007)Le~Guen, Miron, C{\'e}olin, Guillemin,
  Leclercq, Simon, Morin, Mocellin, Bj{\"o}rneholm, Naves~de Brito, and
  Sorensen]{Le-Guen_2007}
K.~Le~Guen, C.~Miron, D.~C{\'e}olin, R.~Guillemin, N.~Leclercq, M.~Simon,
  P.~Morin, A.~Mocellin, O.~Bj{\"o}rneholm, A.~Naves~de Brito and S.~L.
  Sorensen, \emph{The Journal of Chemical Physics}, 2007, \textbf{127},
  114315\relax
\mciteBstWouldAddEndPuncttrue
\mciteSetBstMidEndSepPunct{\mcitedefaultmidpunct}
{\mcitedefaultendpunct}{\mcitedefaultseppunct}\relax
\EndOfBibitem
\bibitem[Hudson \emph{et~al.}(1993)Hudson, Shirley, Domke, Remmers, Puschmann,
  Mandel, Xue, and Kaindl]{Hudson_1993}
E.~Hudson, D.~A. Shirley, M.~Domke, G.~Remmers, A.~Puschmann, T.~Mandel, C.~Xue
  and G.~Kaindl, \emph{Physical Review A}, 1993, \textbf{47}, 361--373\relax
\mciteBstWouldAddEndPuncttrue
\mciteSetBstMidEndSepPunct{\mcitedefaultmidpunct}
{\mcitedefaultendpunct}{\mcitedefaultseppunct}\relax
\EndOfBibitem
\bibitem[Stener \emph{et~al.}(2011)Stener, Bolognesi, Coreno, O'Keeffe, Feyer,
  Fronzoni, Decleva, Avaldi, and Kivim{\"a}ki]{Stener_2011}
M.~Stener, P.~Bolognesi, M.~Coreno, P.~O'Keeffe, V.~Feyer, G.~Fronzoni,
  P.~Decleva, L.~Avaldi and A.~Kivim{\"a}ki, \emph{The Journal of Chemical
  Physics}, 2011, \textbf{134}, 174311\relax
\mciteBstWouldAddEndPuncttrue
\mciteSetBstMidEndSepPunct{\mcitedefaultmidpunct}
{\mcitedefaultendpunct}{\mcitedefaultseppunct}\relax
\EndOfBibitem
\bibitem[Blancard \emph{et~al.}(2012)Blancard, Coss{\'e}, Faussurier, Bizau,
  Cubaynes, El~Hassan, Guilbaud, Al~Shorman, Robert, Liu, Nicolas, and
  Miron]{Blancard_2012}
C.~Blancard, P.~Coss{\'e}, G.~Faussurier, J.~M. Bizau, D.~Cubaynes,
  N.~El~Hassan, S.~Guilbaud, M.~M. Al~Shorman, E.~Robert, X.~J. Liu, C.~Nicolas
  and C.~Miron, \emph{Physical Review A}, 2012, \textbf{85}, 043408--\relax
\mciteBstWouldAddEndPuncttrue
\mciteSetBstMidEndSepPunct{\mcitedefaultmidpunct}
{\mcitedefaultendpunct}{\mcitedefaultseppunct}\relax
\EndOfBibitem
\bibitem[Aksela \emph{et~al.}(1992)Aksela, Aksela, Hotokka, Yagishita, and
  Shigemasa]{H_Aksela_1992b}
H.~Aksela, S.~Aksela, M.~Hotokka, A.~Yagishita and E.~Shigemasa, \emph{Journal
  of Physics B: Atomic, Molecular and Optical Physics}, 1992, \textbf{25},
  3357\relax
\mciteBstWouldAddEndPuncttrue
\mciteSetBstMidEndSepPunct{\mcitedefaultmidpunct}
{\mcitedefaultendpunct}{\mcitedefaultseppunct}\relax
\EndOfBibitem
\bibitem[Kivimaki \emph{et~al.}(1993)Kivimaki, Aksela, Aksela, Yagishita, and
  Shigemasa]{Kivimaki_1993}
A.~Kivimaki, H.~Aksela, S.~Aksela, A.~Yagishita and E.~Shigemasa, \emph{Journal
  of Physics B: Atomic, Molecular and Optical Physics}, 1993, \textbf{26},
  3379\relax
\mciteBstWouldAddEndPuncttrue
\mciteSetBstMidEndSepPunct{\mcitedefaultmidpunct}
{\mcitedefaultendpunct}{\mcitedefaultseppunct}\relax
\EndOfBibitem
\bibitem[Ninomiya(1981)]{Ninomiya_1981}
K.~Ninomiya, \emph{Journal of Physics B: Atomic and Molecular Physics}, 1981,
  \textbf{14}, 1777--1790\relax
\mciteBstWouldAddEndPuncttrue
\mciteSetBstMidEndSepPunct{\mcitedefaultmidpunct}
{\mcitedefaultendpunct}{\mcitedefaultseppunct}\relax
\EndOfBibitem
\bibitem[LANDAU and LIFSHITZ(1976)]{Landau_1976}
L.~D. LANDAU and E.~M. LIFSHITZ, in \emph{MECHANICS}, Butterworth-Heinemann,
  Oxford, 1976, ch. IV - COLLISIONS BETWEEN PARTICLES, pp. 41--57\relax
\mciteBstWouldAddEndPuncttrue
\mciteSetBstMidEndSepPunct{\mcitedefaultmidpunct}
{\mcitedefaultendpunct}{\mcitedefaultseppunct}\relax
\EndOfBibitem
\bibitem[Laskin and Lifshitz(2001)]{Laskin_2001}
J.~Laskin and C.~Lifshitz, \emph{Journal of Mass Spectrometry}, 2001,
  \textbf{36}, 459--478\relax
\mciteBstWouldAddEndPuncttrue
\mciteSetBstMidEndSepPunct{\mcitedefaultmidpunct}
{\mcitedefaultendpunct}{\mcitedefaultseppunct}\relax
\EndOfBibitem
\end{mcitethebibliography}
\bibliographystyle{rsc} 

\end{document}